\documentclass{aa}  
\usepackage{graphicx}
\usepackage{microtype}
\usepackage{placeins}
\usepackage{txfonts}
\usepackage{xspace}
\usepackage{multicol}
\usepackage{multirow}
\usepackage{color}
\usepackage{url}
\usepackage{hyperref}
\usepackage[flushleft]{threeparttable}
\usepackage[switch]{linenoaa} 
\newcommand\nicer{{NICER}\xspace} %not a satellite
\newcommand\nustar{\textit{NuSTAR}\xspace}
\newcommand\ixpe{\textit{IXPE}\xspace}

\usepackage{graphicx}
\usepackage{txfonts}

\begin{document}

\title{Discovery of energy-dependent phase variations in the polarization angle of Cen X-3} 

\titlerunning{X-ray polarization  of Cen X-3}

\author{Qing-Chang Zhao
\inst{\ref{in:IHEP}, \ref{in:UCAS}} 
\and 
Lian Tao 
\inst{\ref{in:IHEP}}\thanks{E-mail: taolian@ihep.ac.cn}
\and 
Sergey S. Tsygankov\inst{\ref{in:UTU},\ref{in:IHEP}} \thanks{E-mail: sergey.tsygankov@utu.fi}
\and 
Juri Poutanen\inst{\ref{in:UTU}} 
\and
Hua Feng 
\inst{\ref{in:IHEP}}
\and 
Shuang-Nan Zhang 
\inst{\ref{in:IHEP},\ref{in:UCAS}}
\and 
Hancheng Li
\inst{\ref{in:Geneva}}
\and 
Mingyu Ge
\inst{\ref{in:IHEP}}
\and
Liang Zhang 
\inst{\ref{in:IHEP}}
\and
Alexander~A.~Mushtukov
\inst{\ref{in:Oxford},\ref{in:UCL}}
}

\institute{
State Key Laboratory of Particle Astrophysics, Institute of High Energy Physics, Chinese Academy of Sciences, Beijing 100049, China \label{in:IHEP}   
\and 
University of Chinese Academy of Sciences, Chinese Academy of Sciences, Beijing 100049, China
\label{in:UCAS}
%\email{email here} 
\and 
Department of Physics and Astronomy, FI-20014 University of Turku,  Finland \label{in:UTU} 
\and 
Department of Astronomy, University of Geneva, 16 Chemin d’Ecogia, Versoix, CH-1290, Switzerland \label{in:Geneva}
\and 
Astrophysics, Department of Physics, University of Oxford, Denys Wilkinson Building, Keble Road, Oxford OX1 3RH, UK \label{in:Oxford}
\and
Mullard Space Science Laboratory, University College London, Holmbury St. Mary, Surrey RH5 6NT, UK  \label{in:UCL}
} 

\abstract{We present a detailed polarimetric analysis of Cen X-3 using \ixpe observations during its high state, revealing complex energy-dependent polarization behavior. While phase-averaged polarization shows marginal energy dependence, phase-resolved analysis reveals that the energy dependence of the polarization angle (PA) is strongly phase-dependent, with dramatic variations visible in a few specific phase intervals. We model this behavior using a two-component polarization framework consisting of a pulsed component governed by the Rotating Vector Model (RVM) and an additional phase-dependent component. By allowing the additional component's polarized flux to vary with pulse phase while fixing its PA, the observed complex behavior can be reconciled with a single set of RVM parameters across all energies. Spectroscopic analysis using \ixpe, \nicer and \nustar during the high state reveals phase-modulated intrinsic hydrogen column density and covering fraction, suggesting that the wind properties are modulated with pulse phase. Our findings indicate that phase-dependent scattering in the disk wind may significantly alter the observed polarization properties of X-ray pulsars.}

\keywords{magnetic fields – methods: observational – polarization – pulsars: individual: Cen X-3 – stars: neutron – X-rays: binaries}

\maketitle
%
%
%-------------------------------------------------------------------

\section{Introduction}
Accreting X-ray pulsars (XRPs) are binary systems harboring strongly magnetized neutron stars, with the dipole magnetic fields reaching $10^{12}$--$10^{13}$~G, accreting material from a donor star. The interaction between strong magnetic fields, radiation, and accreting matter leads to the diverse and complex observational behavior of XRPs \citep[see][for a recent review]{Mushtukov_Tsygankov_review}. The strong magnetic fields fundamentally alter the underlying physical processes, for example, the Compton scattering cross-section. Due to the significant difference in opacity between the ordinary (O) and extraordinary (X) modes in the highly magnetized plasma of XRPs, radiation was previously expected to be strongly polarized, with the polarization degree (PD) reaching $\sim$80\% \citep{Meszaros1988,Caiazzo_polarization_model}. 
The launch of the Imaging X-ray Polarimetry Explorer  \citep[\ixpe;][]{Soffitta_etal_2021,Weisskopf_ixpe} in December 2021 has opened a new window for testing the X-ray polarization properties predicted by theoretical models.

To date, more than a dozen XRPs have  been observed by \ixpe, including \mbox{Her~X-1} \citep{Doroshenko_etal_2022_herx-1,Garg_herx-1,Zhao_herx-1,Heyl_herX-1}, \mbox{Cen~X-3} \citep{Tsygankov_etal_2022_cenx-3}, GRO~J1008$-$57 \citep{Tsygankov_groj1008}, 4U~1626$-$67 \citep{Marshall_4U1627}, X~Persei \citep{Mushtukov_xpeisei}, \mbox{Vela~X-1} \citep{Forsblom_velaX-1,Forsblom2025}, EXO~2030+375 \citep{Malacaria_exo2030}, GX~301$-$2 \citep{Suleimanov_gx301}, RX~J0440.9+44331 / LS~V~+44~17 \citep{Doroshenko_etal_2023,Zhao_RXJ0440}, Swift~J0243.6+6124 \citep{SwiftJ0243_Majumder,SwiftJ0243_Poutanen}, \mbox{SMC~X-1} \citep{SMCX-1_Forsblom}, and 4U~1538$-$52 \citep{Loktev25}. Surprisingly, all of these sources exhibit PDs that are significantly below theoretical predictions, even in phase-resolved measurements. Most \ixpe targets were observed below the critical luminosity at which the accretion geometry transitions between a surface hot spot and an accretion column \citep{Basko_etal_1976}. Only a few sources, including RX~J0440.9+4431 / LS~V~+44~17 and SMC~X-1, were observed in the super-critical regime, but they also exhibited low PDs. Another source, 1A~0535+262, observed by PolarLight in the super-critical regime, showed no significant polarization with a 99\% confidence upper limit of 34\% in the 3--8~keV band \citep{Polarlight, Long_0535}.

Although the PD often shows erratic variations over the pulse phase, the polarization angle (PA) can, in most cases, be well modeled using the rotating vector model \citep[RVM;][]{Radhakrishnan_RVM,Meszaros1988,Poutanen_RVM}. This behavior aligns with the predictions of vacuum birefringence \citep{Gnedin78,Pavlov79}, where the photon’s polarization direction is expected to follow the local magnetic field geometry until it decouples at the adiabatic radius \citep{Heyl00,Heyl02PRD,Taverna15}. 
This radius is about 20 neutron star radii for keV  photons and the typical surface magnetic field strength detected in XRPs \citep{Heyl18,Taverna24}. 
At such a distance,  the magnetic field is predominantly dipolar, resulting in the observed PA being either parallel or perpendicular to the instantaneous projection of the magnetic dipole axis onto the plane of the sky, depending on the dominant intrinsic polarization mode (O-mode or X-mode). Consequently, the phase dependence of the PA is a purely geometrical effect, and fitting the RVM to PA variations across the pulse phase provides a unique tool for constraining the geometry of XRPs.
However, in RX~J0440.9+4431 / LS~V~+44~17, the RVM parameters were found to vary dramatically between two observations separated by only $\sim$20~d, a timescale that is difficult to reconcile with precession models. 
To address this discrepancy, an additional polarized component, assumed to be constant across the pulse phase, has been proposed \citep{Doroshenko_etal_2023}. A similar component has also been reported in Swift~J0243.6+6124 \citep{SwiftJ0243_Poutanen}. After subtracting this additional component, the PA variations of the pulsar component across different observations can be well described by the RVM with the same set of parameters.

With \ixpe, it is now possible to investigate how the polarization properties of XRPs vary with energy, although this is limited to the relatively narrow 2--8\,keV band. In X~Persei, a remarkable increase in PD with energy has been observed -- rising from nearly zero at 2~keV to about 30\% at 8~keV \citep{Mushtukov_xpeisei}; yet the physical mechanism driving this trend remains unclear. In \mbox{Vela~X-1}, \citet{Forsblom2025} reported a $90\degr$ swing in the PA between low (2--3~keV) and high (5--8~keV) energies, which may arise from two spectral components featuring different PAs or from the vacuum resonance. More recently, \citet{Loktev25} reported a $70\degr$ shift in the pulse phase-averaged PA between low and high energies in 4U~1538$-$52, and pulse phase-resolved analysis also revealed distinct behavior in these separate energy ranges. However, the underlying physical reason for energy-dependent behavior in this source remains unknown.

%%%%%%%%%%%%%%%%%%%%%%%%%%%%%%%%%%%
%\begin{figure*}%[h!]
%\centering
%\includegraphics[width=0.8\textwidth]%{fig/Lightcurve.pdf}
%\caption{\ixpe light curves of Cen X-3 in 2--8 keV energy band with a 1000\,s binsize. }
%\label{fig:lc}
%\end{figure*}
%%%%%%%%%%%%%%%%%%%%%%%%%%%%%%%%%%%

%--------------------------------------------------------------------
\begin{table*}%[ht]
\centering
\caption{Log of the observations used in this paper. }
\begin{tabular}{lccc}
\hline\hline
Instrument & ObsID & Start time (UTC) & Exposure (ks)\\
\hline
 \ixpe & 01250201 & 2022-07-04 06:27:52 & 178.8 (DU1) / 178.9 (DU2) / 178.9 (DU3) \\
\nustar & 30901020002 &  2024-02-13 21:21:09  & 23.0 (FPMA) / 23.5 (FPMB) \\
\nicer & 7614010104 &  2024-02-14 00:53:44  & 4.3 \\
\hline
\end{tabular}
\label{tab:log}
\end{table*}

%%%%%%%%%%%%%%%%%%%
\begin{figure*}%[h!]
\centering
\includegraphics[width=1.0\textwidth]{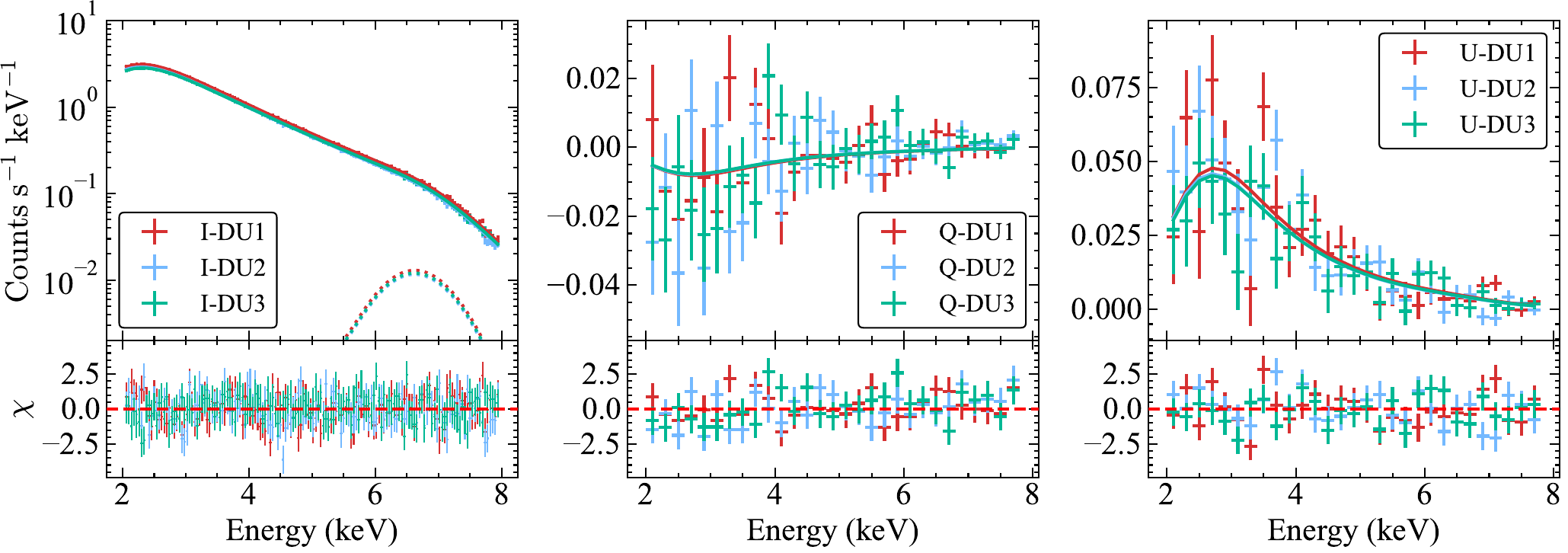}
\caption{Energy distributions of the Stokes parameters $I$, $Q$, and $U$ and the best-fit model (top panels) and the residuals between data and model normalized for the errors are shown in the bottom panels.} 
\label{fig:ixpe_spec_fit}
\end{figure*}
%%%%%%%%%%%%%%%%%%%%
\mbox{Cen~X-3}, the first discovered accreting X-ray pulsar, has a spin period of $P_{\rm spin} \approx 4.8\,\mathrm{s}$ \citep{Giacconi1971}. It is an eclipsing high-mass X-ray binary consisting of a neutron star with a mass of $1.34^{+0.16}_{-0.14}~M_{\odot}$ in an almost circular orbit around an O6--8\,II-III supergiant V779~Cen, which has a mass of $M_{\rm O} = 20.5 \pm 0.7\,M_{\odot}$ \citep{Krzeminski1974,Ash1999,Raichur2010}. The orbital period of the system is $\sim$2.08\,d, and during about 20\% of the orbit the pulsar is eclipsed by the companion star owing to a high inclination of $70\fdg2 \pm 2\fdg7$ \citep{Ash1999}. The distance to \mbox{Cen~X-3} is $D = 6.4^{+1.0}_{-1.4}\,\mathrm{kpc}$, derived from \textit{Gaia} parallax measurements \citep{Arnason_Gaia}.

\citet{Tsygankov_etal_2022_cenx-3} performed a detailed polarimetric analysis of \mbox{Cen~X-3} using two \ixpe datasets obtained during low- and high-luminosity states. In the high-luminosity state, a highly significant polarization (at $\sim 20\sigma$) was detected. By applying the RVM to the pulse-phase variations in the PA, the geometrical parameters were rather well constrained. In this paper, we further explore the energy-dependent polarimetric properties of the source. The paper is organized as follows. 
In Sect.~\ref{sec:sec2}, we describe the observations and data reduction methods.
The results are presented in Sect.~\ref{sec:sec3} and discussed in Sect.~\ref{sec:sec4}.

%cen x-3的简单介绍，包括前面sergey paper的结果。
%文章的组织。

\section{Observations and data reduction} \label{sec:sec2}

\subsection{\ixpe}

\ixpe is a joint NASA--Italian Space Agency (ASI) mission. The observatory, equipped with three grazing-incidence X-ray telescopes and gas pixel detectors, is dedicated to performing polarimetric measurements in the 2--8 keV energy range \citep{Soffitta_etal_2021,Weisskopf_ixpe}. \ixpe has so far performed two observations of Cen X-3: the first on January 29--31, 2022, when the source was in the low state, and the second on July 4--7, 2022, during a high state \citep{Tsygankov_etal_2022_cenx-3}.  
In this work, we focus on the high-state observation (see Table~\ref{tab:log}), which exhibits a statistically significant polarization. The data were processed using \textsc{ixpeobssim} version 31.0.3 \citep{Baldini_etal_2022}.\footnote{\url{https://ixpeobssim.readthedocs.io/en/latest/}} For the analysis, a circular extraction region with a radius of 90\arcsec\ centered on the source was used. Background subtraction was not applied, as its contribution was deemed negligible \citep{Di_Marco_etal_2023}.  
Photon arrival times were corrected to the Solar System barycenter using the \texttt{barycorr} tool from the \textsc{HEASoft} package (v6.36.0). Binary orbital modulation was corrected using the ephemeris and orbital parameters provided by \textit{Fermi}/GBM.\footnote{\url{https://gammaray.msfc.nasa.gov/gbm/science/pulsars/lightcurves/cenx3.html}} 
We used \texttt{xselect} to extract weighted Stokes $I$, $Q$, and $U$ spectra using the command ``extract SPECTRUM stokes=NEEF'' \citep{DiMarco_weighted2022}, and employed \texttt{ixpecalcarf} to generate the corresponding ARF/MRF files with weights set to 1. For spectral analysis, the Stokes $I$ spectra were grouped to ensure a minimum of 25 counts per bin, while the Stokes $Q$ and $U$ spectra were binned with a constant energy interval of 0.2~keV. The best-fit parameters were determined by minimizing $\chi^2$ statistics. Additionally, we generated polarization cubes using the \texttt{pcube} algorithm to perform model-independent polarization analysis.
%The \texttt{v12} version of the weighted response files was used to generate the polarization and spectral products.  We generated polarization cubes with the \texttt{pcube} algorithm and the Stokes \textit{I}, \textit{Q}, and \textit{U} spectra using the \texttt{PHA1}, \texttt{PHA1Q}, and \texttt{PHA1U} algorithms, respectively. 
Throughout this paper, uncertainties are reported at a 68\% confidence level unless otherwise stated.

\subsection{\nustar}
The Nuclear Spectroscopic Telescope Array \citep[\nustar;][]{Harrison_nustar} has conducted several observations of \mbox{Cen~X-3}. Without strictly simultaneous \ixpe coverage, we selected an archival \nustar observation in a high-flux state (ObsID: 30901020002; see Table~\ref{tab:log}) with flux levels comparable to the \ixpe observations for our broadband spectral analysis.

Data reduction was carried out using the \texttt{nupipeline} routine from the \texttt{NuSTARDAS} package within \textsc{HEASoft} v6.36.0, along with the calibration database (CALDB) version 20250729. Barycentric and binary orbital corrections were applied. Source and background events were extracted from circular and annular regions with radii of 90\arcsec\ and 120\arcsec--150\arcsec, respectively. Spectra were generated using \texttt{nuproducts}, grouped with a minimum of 25 counts per bin. We use the 4–79~keV range for \nustar\ and omit 3–4~keV due to its inconsistency with the \nicer\ spectrum.

\subsection{\nicer}

%The Neutron Star Interior Composition Explorer \citep[{\nicer};][]{Gendreau_NICER} is a soft X-ray instrument mounted on the International Space Station. It consists of 56 co-aligned concentrator X-ray optics, each paired with a single-pixel silicon drift detector, enabling high-precision timing and spectroscopy in the soft X-ray band.  

The Neutron Star Interior Composition Explorer \citep[{\nicer};][]{Gendreau_NICER} enables high-precision timing and spectroscopy in the soft X-ray band. To complement the low-energy coverage, we selected a {\nicer} observation quasi-simultaneous with the \nustar one (see Table~\ref{tab:log}). Data reduction was performed using \texttt{nicerl2} with \texttt{underonly\_range="0-200"} and \texttt{overonly\_range="0-2"} with the CALDB version 20240206. Focal Plane Modules (FPMs) 14 and 34 were excluded due to elevated background noise. Light curves and spectra were extracted using \texttt{nicerl3-lc} and \texttt{nicerl3-spec}, respectively. The SCORPEON model was employed to estimate the background. The spectra were grouped to ensure a minimum of 25 counts per bin, and we restricted our spectral analysis to the 0.7--10.0 keV energy range.

%%%%%%%%%%%%%%%%%%%%%%%%%%%%%%%%%%%

%%%%%%%%%%%%%%%%%%%%%%%%%%%%%%%%%%%
\begin{table*}%[ht]
\centering
\caption{The phase-averaged spectro-polarimetric results with \texttt{constant*tbabs*polconst*(powerlaw+gauss)} and \texttt{constant*tbabs*pollin*(powerlaw+gauss)}. The first and second columns list the best-fit values obtained from the \texttt{polconst} and \texttt{pollin} models, respectively, while the third column shows the \texttt{pollin} model with the PD slope fixed at zero. }
\begin{tabular}{lcccc}
\hline\hline 
Model & Parameter &Value & Value & Value \\
\hline
 \texttt{tbabs} & $N_{\rm H}$ ($10^{22} \ \rm cm^{-2}$) & $2.45\pm0.04$ & $2.45\pm0.04$ & $2.45\pm0.04$\\
\texttt{polconst} & PD (\%) & $5.6\pm0.3$ & - & -\\
        & PA (deg) & $49.9\pm1.4$& - & -\\
\texttt{pollin} & PD (\%) & - & $6.5\pm0.7$ & $5.7\pm0.3$ \\
&PD slope (\% $\rm keV^{-1}$)& - & $-0.3\pm0.2$ &$0^{\mathrm{fixed}}$\\
        & PA (deg) & - &$61\pm4$ & $61\pm4$\\
        & PA slope ($\rm deg \ keV^{-1}$)& - &$-3.9\pm1.2$ & $-3.9\pm1.2$\\
\texttt{powerlaw} &  $\Gamma$ &$1.17\pm0.01$  & $1.17\pm0.01$ & $1.17\pm0.01$ \\
 &   Normalization & $0.64\pm0.01$ & $0.64\pm0.01$ & $0.64\pm0.01$\\
\texttt{gauss} &  Line energy (keV) &  $6.75^{+0.05}_{-0.04}$ &$6.75^{+0.05}_{-0.04}$  &  $6.75^{+0.05}_{-0.04}$\\
 &  Sigma (keV) & $0.33\pm0.08$  &$0.33\pm0.08$ & $0.33\pm0.08$ \\
 &  Normalization & $0.010\pm0.001$ & $0.010\pm0.001$ &  $0.010\pm0.001$\\
 \hline
 \texttt{constant} & DU1 & $1^{\mathrm{fixed}}$ &$1^{\mathrm{fixed}}$  & $1^{\mathrm{fixed}}$\\
 & DU2 & $1.014\pm0.002$ & $1.014\pm0.002$  & $1.014\pm0.002$\\
 & DU3 & $0.986\pm0.002$ & $0.986\pm0.002$  & $0.986\pm0.002$\\ \hline
 
& $\chi^2$/dof & 595.8/608 & 583.4/606 & 585.0/607 \\  
\hline
\end{tabular}
\label{tab:ixpe_spec_fit}
\end{table*}

\section{Results and analysis} \label{sec:sec3} 

In this paper, time intervals during eclipse ingress and egress were excluded from subsequent analysis to avoid contamination.
%In Fig.~\ref{fig:lc}, we present the 2--8 keV \ixpe light curve of \mbox{Cen~X-3}. During this observation, the source was in the high state \citep{Tsygankov_etal_2022_cenx-3}. A sharp flux drop, caused by the eclipse of the pulsar by its companion, is clearly visible. Time intervals during eclipse ingress and egress were excluded from subsequent analysis to avoid contamination.
%%%%%%%%%%%%%%%%%%%%%%%%%%%%%%%%%%%

%%%%%%%%%%%%%%%%%%%%%%%%%%%%%%%%%%%

%%%%%%%%%%%%%%%%%%%%%%%%%%%%%%%%%%%
\begin{figure}%[h!]
\centering
\includegraphics[width=0.90\linewidth]{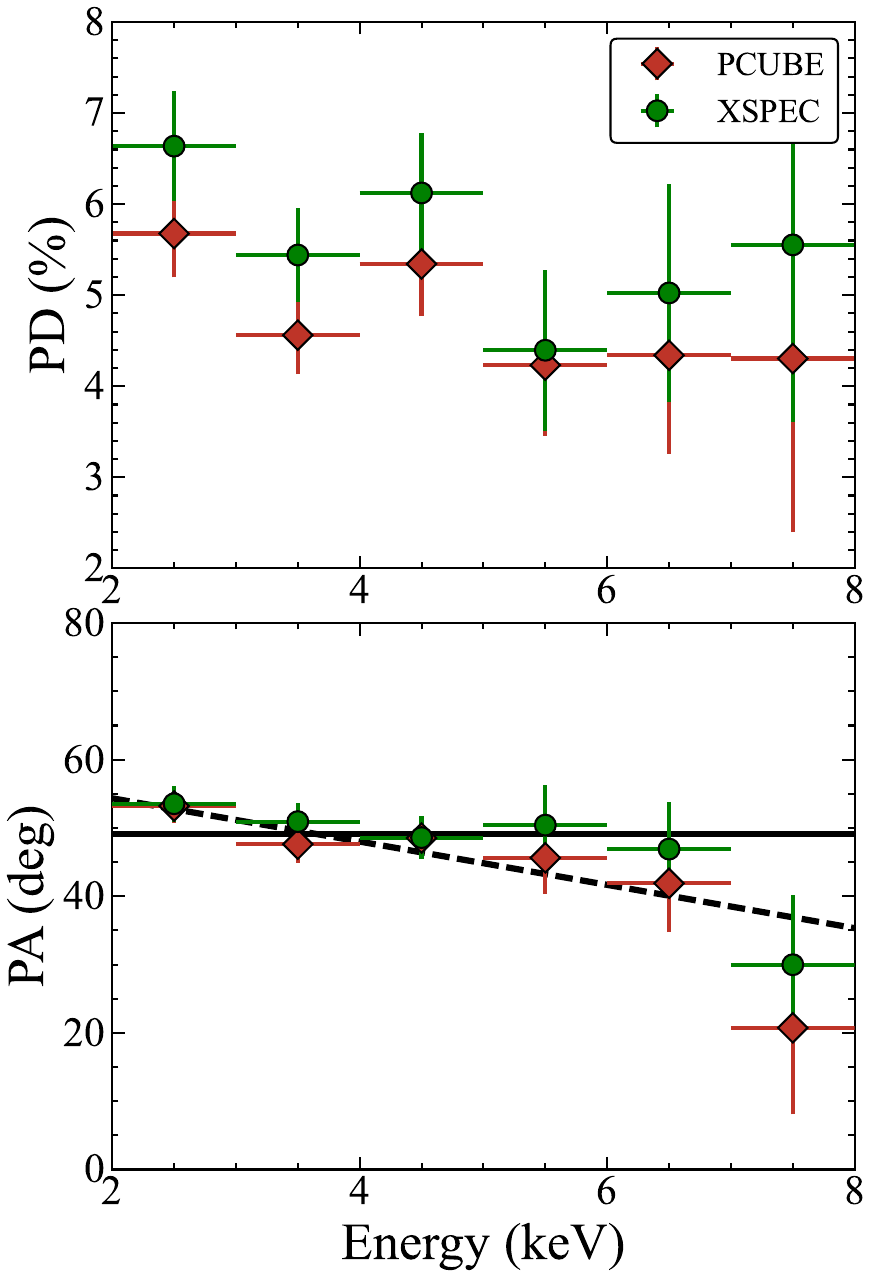}
\caption{Phase-averaged energy dependence of the PD and PA for \mbox{Cen~X-3}. The black solid and dashed lines represent the constant and linear fits to the PA variation with energy, obtained using maximum likelihood estimation. 
%The Bayesian information criterion (BIC) values for the constant and linear models are 17.37 and 6.88, respectively.} 
}
\label{fig:energy_dependence_aver}
\end{figure}
%%%%%%%%%%%%%%%%%%%%%%%%%%%%%%%%%%%

\subsection{Polarimetric analysis}
We first performed a pulse phase-averaged, model-independent polarimetric analysis \citep{Kislat2015} using the \texttt{pcube} algorithm within \texttt{XPBIN}. This yielded a PD of $4.7\%\pm0.4\%$ and a PA of $46\fdg6\pm2\fdg0$ in 2--8 keV.
We then conducted a spectro-polarimetric analysis by jointly fitting the Stokes \textit{I}, \textit{Q}, and \textit{U} spectra using \textsc{xspec} \citep{Arnaud_xspec}. We first adopted the \texttt{constant*polconst*tbabs*powerlaw} model, which provided a poor fit with $\chi^2/\mathrm{dof} = 785.8/611$. The residuals are obvious in 6--8 keV, so we add a \texttt{gaussian} to fit this feature, which significantly improved the fit, yielding $\chi^2/\mathrm{dof} = 595.8/608$. The data and best-fit model are shown in  Fig.~\ref{fig:ixpe_spec_fit}. This spectro-polarimetric fit resulted in a PD of $5.6\%\pm0.3\%$ and a PA of $49\fdg9\pm1\fdg4$, as listed in Table~\ref{tab:ixpe_spec_fit}, consistent with the model-independent analysis. The PD and PA derived from the spectro-polarimetric analysis using the \texttt{polconst} model are in good agreement with the results reported by \citet{Tsygankov_etal_2022_cenx-3}. The spectral parameter of photon index, $\Gamma$, is slightly lower than in \citet{Tsygankov_etal_2022_cenx-3}. We note that \citet{Tsygankov_etal_2022_cenx-3} performed their analysis over the 2--7~keV energy range, whereas we used a broader band (2--8~keV). Additionally, our analysis includes a correction for vignetting effects using the \texttt{ixpecalcarf} tool.

%Replacing the \texttt{powerlaw} with a broken power-law model (\texttt{Bknpower}) significantly improved the fit, yielding $\chi^2/\mathrm{dof} = 649.19/609$. We note that the use of \texttt{Bknpower} is phenomenological.
%This spectro-polarimetric fit resulted in a PD of $5.6\%\pm0.3\%$ and a PA of $49.9\degr\pm1.4\degr$, consistent with the model-independent analysis.
\begin{figure*}%[h!]
\centering
\includegraphics[width=0.99\textwidth]{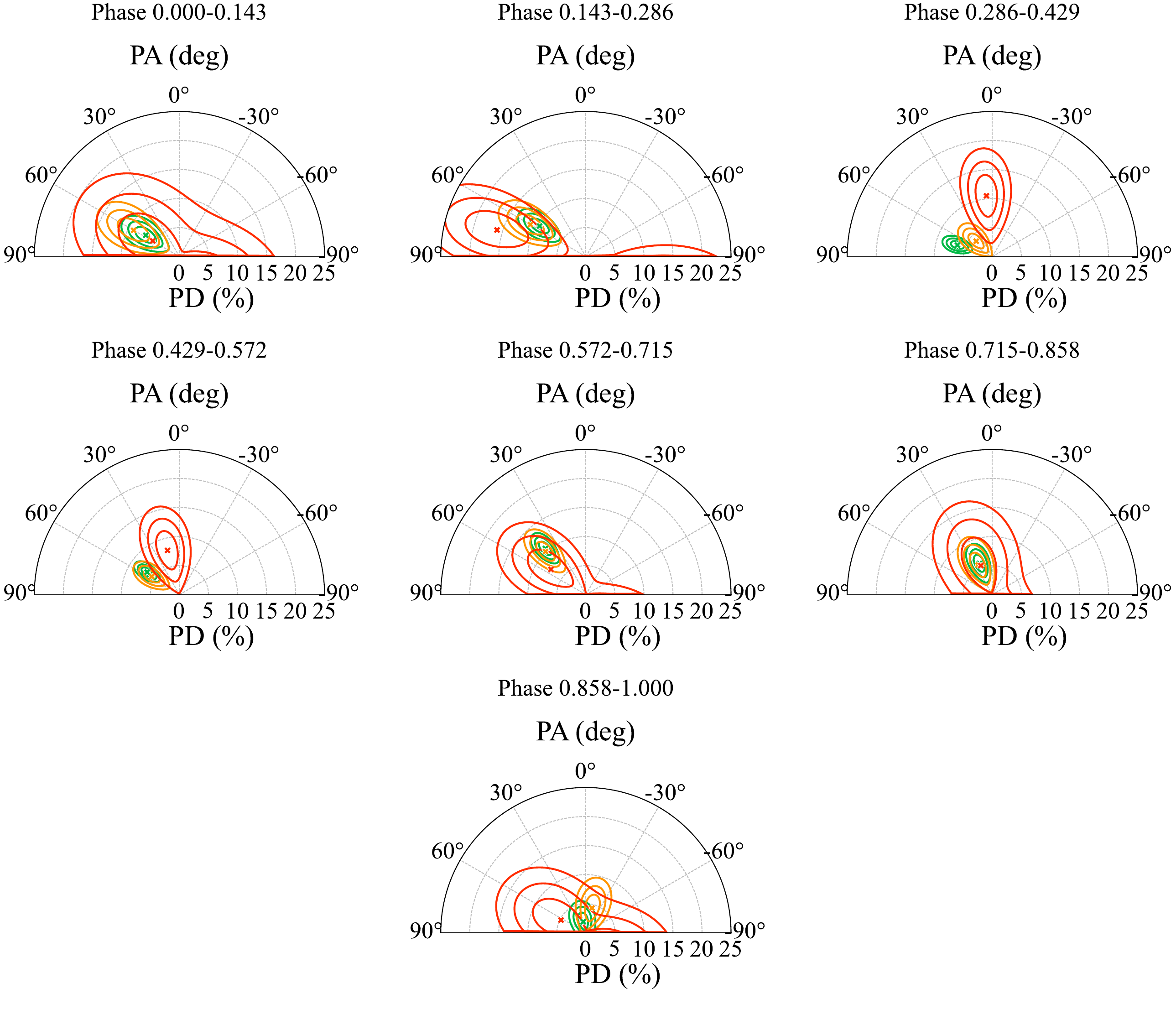}

\caption{Polarization vectors as a function of pulse phase for Cen X--3 in three energy bands: 2--4 keV (green), 4--6 keV (orange), and 6--8 keV (red). In each plot, the PD and PA contours are shown at the 68.27\%, 95.45\%, and 99.73\% confidence levels.}

\label{fig:phase_energy}
\end{figure*}
%%%%%%%%%%%%%%%%%%%%%%%%%%%%%%%%%%%
\begin{figure*} %[h!]
\centering
\includegraphics[width=0.9\textwidth]{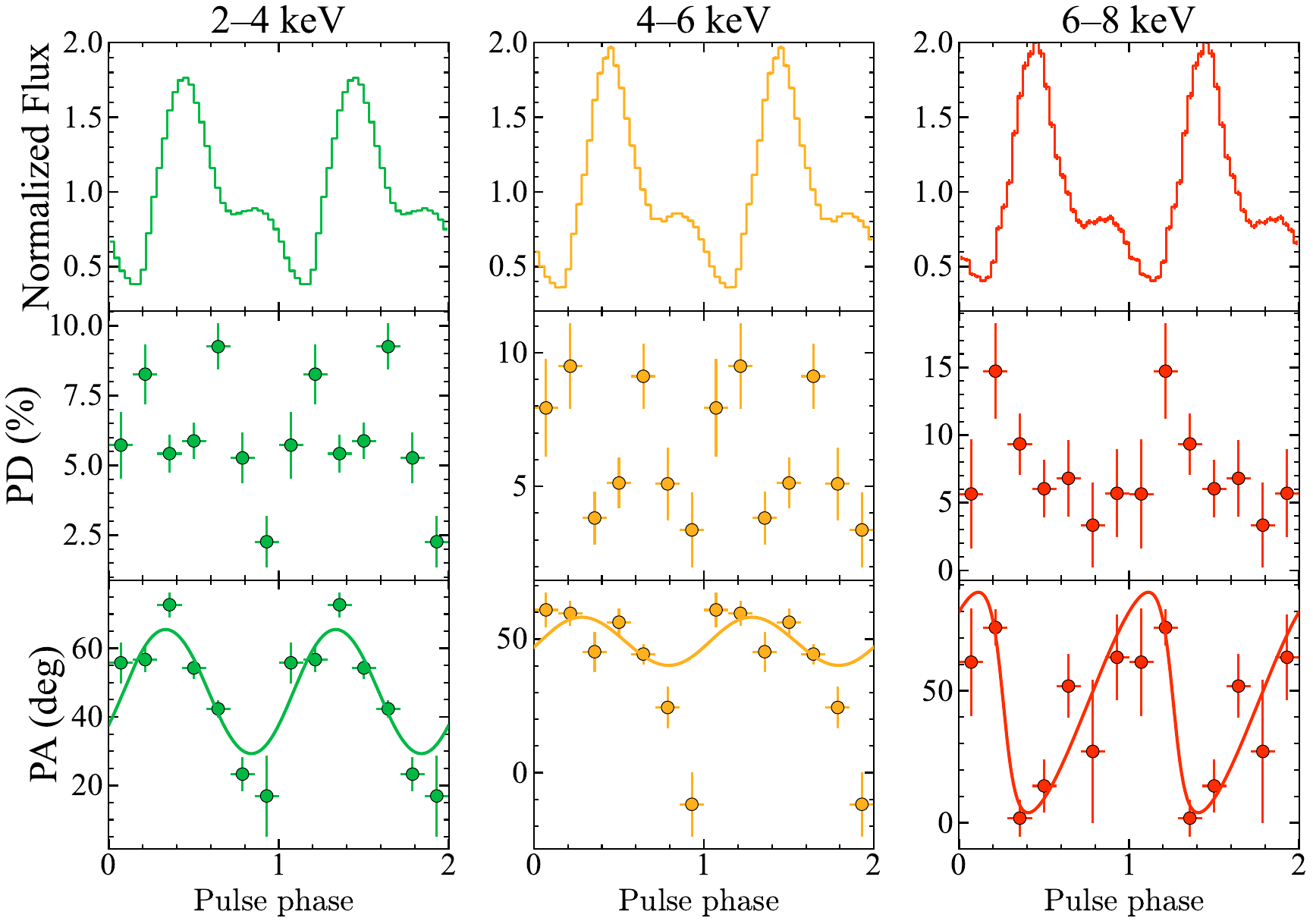}
\caption{Pulse phase--resolved, energy-dependent polarimetric results. Top panels: normalized pulse profiles in three energy bands. Middle panels: variations of PD with pulse phase. Bottom panels: variations of PA with pulse phase. The 2--4, 4--6, and 6--8 keV bands are shown in the left, middle, and right columns, respectively, and are color-coded in green, orange, and red. The solid curves in the bottom panels represent the best-fitting single-component RVM models applied to the PA data.}
\label{fig:phase_resolved}
\end{figure*}
%%%%%%%%%%%%%%%%%%%%%%%%%%%%%%%%%%%

\begin{figure} %[h!]
\centering
\includegraphics[width=0.95\linewidth]{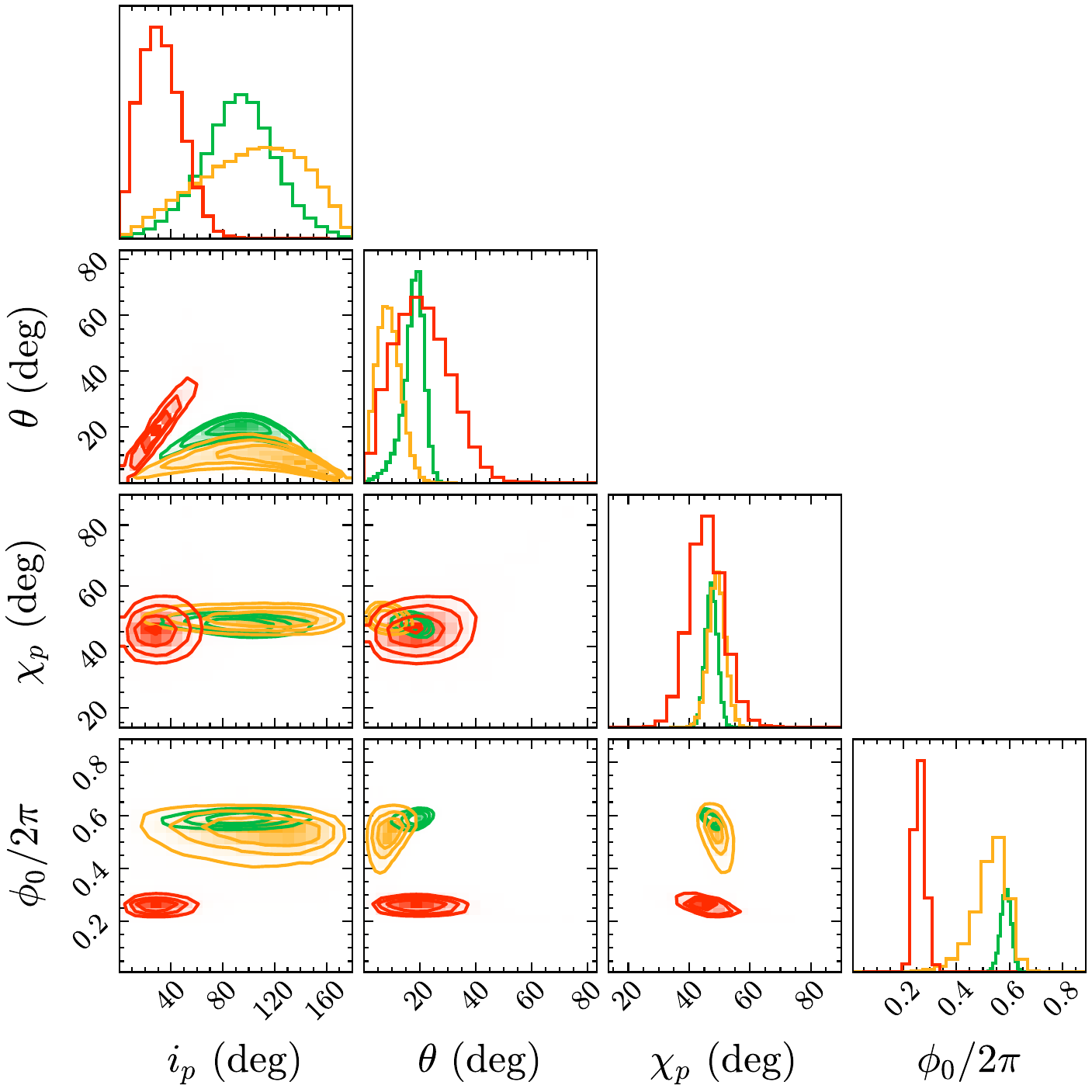}
\caption{Corner plots of the posterior distributions for the single-component RVM parameters for 2--4 keV (green), 4--6 keV (orange) and 6--8 keV (red), respectively. }
\label{fig:corner_plot_rvm}
\end{figure}

To test for a possible energy dependence of the polarization properties in Cen X-3, we replaced the \texttt{polconst} component in the best-fit model with \texttt{pollin}. Applying this modified model to the phase-averaged data resulted in a modest improvement in fit quality, giving $\chi^2/\mathrm{dof} = 583.4/606$, a PA slope of $-3.9 \pm 1.2\,\mbox{deg}\,\mbox{keV}^{-1}$, and a PD slope of $-0.3 \pm 0.2\%\,\mbox{keV}^{-1}$, as listed in Table~\ref{tab:ixpe_spec_fit}. An F-test yielded a $p$-value of 0.0017, indicating a potential energy dependence of the polarization. The slope of the PD is close to zero, indicating that the PD is approximately constant with energy. Accordingly, we fixed the PD slope at zero, which resulted in a PA value and PA slope identical to those obtained when the PD slope was allowed to vary. We further evaluated this using maximum likelihood estimation (MLE) for both constant and linear PA models. The Bayesian information criterion (BIC) values for the constant and linear models were 11.46 and 6.51, respectively, suggesting that the linear model is preferred. This preference for a linear energy dependence of the PA is consistent with the findings of \citet{Tsygankov_etal_2022_cenx-3}.

We performed an energy-resolved analysis using \texttt{pcube}, dividing the 2--8 keV range into six 1-keV wide sub-bands, with results shown in Fig.~\ref{fig:energy_dependence_aver}.
We conducted a complementary energy-resolved spectro-polarimetric analysis within individual sub-bands. Due to limited photon statistics in the narrow energy bins, spectral parameters in each sub-band were fixed to the best-fit values obtained for the full 2--8 keV range. The \texttt{polconst} model was used to estimate the PD and PA, with only these parameters allowed to vary.
As shown in Fig.~\ref{fig:energy_dependence_aver}, the PD shows no significant energy dependence, while the PA exhibits only a marginal trend with energy.

Given the strong phase dependence of polarization in XRPs, we further explore the  dependence of polarization on both energy and pulse phase, dividing the pulse phase into 7 equal bins. The 2--8 keV band was divided into three 2-keV wide sub-bands, and the results are presented in Figs.~\ref{fig:phase_energy} and ~\ref{fig:phase_resolved}. As shown in the top panel of Fig.~\ref{fig:phase_resolved}, the normalized pulse profiles are largely consistent across energies, suggesting limited spectral evolution over phase in the narrow energy band of \ixpe. %However, 
In terms of polarization properties, we observe that the PA exhibits significant energy dependence in a few phase bins.
%The PA exhibits a simple evolution pattern with pulse phase, exhibiting significant energy dependence. 

We use the RVM to model the phase-resolved PA modulation. 
This model has been widely used to infer geometrical parameters of accreting XRPs from phase-resolved polarimetric data \citep[e.g.,][]{Doroshenko_etal_2022_herx-1, Tsygankov_etal_2022_cenx-3, Mushtukov_xpeisei, Marshall_4U1627, Heyl_herX-1, SMCX-1_Forsblom, Forsblom2025,SwiftJ0243_Poutanen, Zhao_herx-1,Zhao_RXJ0440}.
RVM describes the PA variation as a function of pulse phase under the assumption that radiation escapes predominantly in the O-mode. 
The PA predicted by the RVM is given by \citep{Poutanen_RVM} :
\begin{equation}
\label{equ:RVM}
\tan(\mathrm{PA} - \chi_{\mathrm{p}}) = \frac{-\sin\theta\,\sin(\phi - \phi_{0})}{\sin i_{\mathrm{p}}\,\cos\theta - \cos i_{\mathrm{p}}\,\sin\theta\,\cos(\phi - \phi_{0})},
\end{equation}
where \(i_{\mathrm{p}}\) is the inclination angle between the observer's line of sight and the pulsar's spin axis, \(\theta\) is the magnetic obliquity, \(\chi_{\mathrm{p}}\) is the position angle of the spin axis on the sky, and \(\phi_0\) is the pulse phase at which the magnetic axis is closest to the observer.

%%%%%%%%%%%%%%%%%%%%%%%%%%%%%%%%%%%
\begin{figure}%[h!]
\centering
\includegraphics[width=0.90\linewidth]{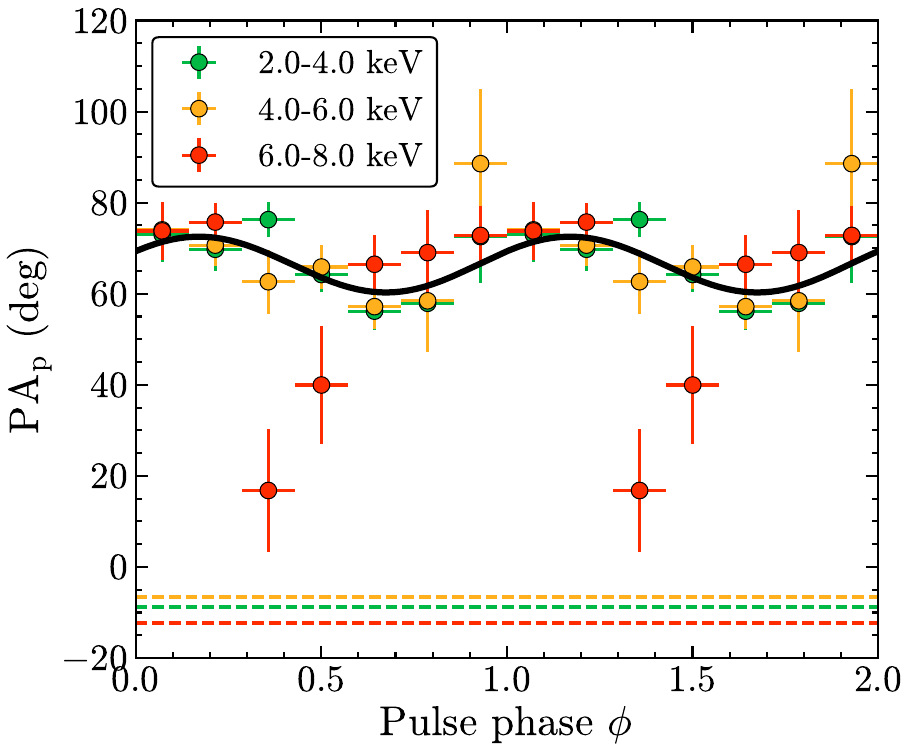}
\caption{PA for the pulsed components of Cen X-3 in different energy bands as indicated by colored points, while the constant components are represented by dashed lines. The black solid line represents the joint RVM fitting curve for three energy bands after subtracting the constant components. The color coding is the same as in Fig.~\ref{fig:phase_resolved}.}
\label{fig:two-comp-fit}
\end{figure}
%%%%%%%%%%%%%%%%%%%%%%%%%%%%%%%%%%%

\begin{figure} %[h!]
\centering
\includegraphics[width=0.9\linewidth]{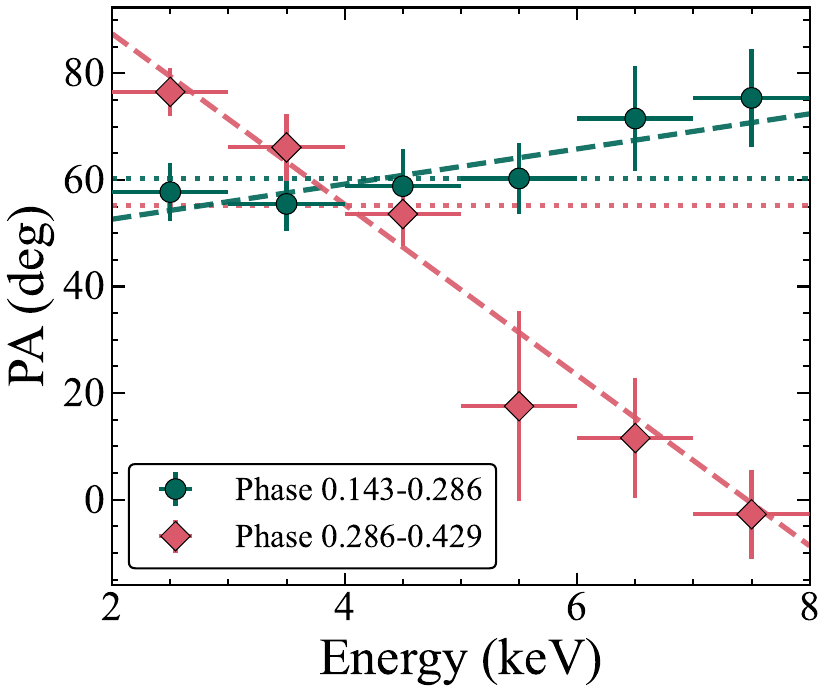}
\caption{Energy dependence of the PA for phase intervals 0.143--0.286 and 0.286--0.429. The solid and dashed lines represent constant and linear fits, respectively. BIC values (constant / linear) are 6.96 / 5.01 for phase 0.143--0.286 and 93.35 / 6.06 for phase 0.286--0.429.}
\label{fig:two_phase_energy}
\end{figure}
%%%%%%%%%%%%%%%%%%%%%%%%%%%%%%%%%%%

%%%%%%%%%%%%%%%%%%%%%%%%%%%%%%%%%%%

To account for the non-Gaussian nature of the PA, we adopted the likelihood formalism introduced by \citet{Naghizadeh1993}, which is based on the probability distribution of the measured PA (denoted $\psi$). The corresponding probability density function \(G(\psi)\) is defined as:

\begin{table} %[h!]
\centering
\caption{Single-RVM best-fit  parameters for three energy bands.}
\begin{tabular}{cccc} \hline\hline
 & \multicolumn{3}{c}{Energy band (keV)} \\ 
 \cline{2-4}
 %\hline
Model & 2.0--4.0 & 4.0--6.0 & 6.0--8.0 \\ \hline
$i_{\rm p}$ (deg)  & $93^{+27}_{-29}$ & $104^{+38}_{-46}$ & $31^{+18}_{-15}$ \\ 
$\theta$ (deg)   & $18.2^{+3.1}_{-4.5}$ & $8.7^{+4.5}_{-4.1}$ & $20^{+11}_{-10}$\\
$\chi_{\rm p}$ (deg)   & $47\pm2$ & $49\pm3$ & $45^{+6}_{-5}$\\
$\phi_0$/2$\pi$   & $0.59\pm0.02$ & $0.54^{+0.05}_{-0.07}$ & $0.26\pm0.02$\\
\hline
\end{tabular}%
   \label{tab:RVM fitting}
\end{table}
\begin{equation}
\label{eq:PA_G}
G(\psi) = \frac{1}{\sqrt{\pi}} 
\left\{ \frac{1}{\sqrt{\pi}} + 
\eta\,\mathrm{e}^{\eta^2} 
\left[1 + \mathrm{erf}(\eta)\right]
\right\} \mathrm{e}^{-p_0^2/2},
\end{equation}
where \(p_0 = \sqrt{q^2 + u^2}/\sigma_{\mathrm{p}}\) is the signal-to-noise ratio of the PD, \(\eta = p_0 \cos[2(\chi - \chi_0)]/\sqrt{2}\), and \(\chi_0 = \frac{1}{2} \arctan(u/q)\) is the measured PA derived from the Stokes parameters. Here, \(\chi\) denotes the model-predicted PA and \(\mathrm{erf}\) is the standard error function. We performed the parameter inference in each energy band using the affine-invariant Markov Chain Monte Carlo (MCMC) sampler \texttt{emcee} \citep{Foreman-Mackey_emcee}. The best-fit parameters from the RVM modeling are summarized in Table~\ref{tab:RVM fitting} with the best-fit models shown at the lower panel of Fig.~\ref{fig:phase_resolved}, and the posterior distributions are shown in Fig.~\ref{fig:corner_plot_rvm}. The position angle of the spin axis (\(\chi_{\rm p}\)) and the magnetic obliquity (\(\theta\)) are broadly consistent across different energy bands. The most notable variations are observed in $i_{\rm p}$ and \(\phi_0\), which show a clear energy dependence. The RVM parameters obtained from the 2–4 keV band closely match those reported by \citet{Tsygankov_etal_2022_cenx-3}, which is not surprising since the low-energy photons tend to dominate when multiple energy bands are combined.

Such variation in RVM parameters across energy bands is generally not expected under standard RVM assumptions. Recent studies of XRPs, notably RX~J0440.9+4431 \citep{Doroshenko_etal_2023,Zhao_RXJ0440} and Swift~J0243.6+6124 \citep{SwiftJ0243_Poutanen}, have reconciled similar discrepancies where RVM parameters appeared to vary significantly over time. This was achieved by introducing an additional phase-independent polarized component.
In this two-component model, the observed emission is modeled as a combination of a pulsed, phase-dependent component and an additional unpulsed, phase-independent one. Under this framework, the PA variations can be reproduced using a single set of RVM parameters. The total Stokes parameters can be expressed as the sum of contributions from both components:
\begin{eqnarray}  
\label{eq:two-comp}
I(\phi) &=& I_{\mathrm{a}} + I_{\mathrm{p}}(\phi), \nonumber \\
Q(\phi) &=& Q_{\mathrm{a}} + Q_{\mathrm{p}}(\phi), \\
U(\phi) &=& U_{\mathrm{a}} + U_{\mathrm{p}}(\phi), \nonumber
\end{eqnarray} 
where the subscripts `p' and `a' denote the pulsed and additional polarized components, respectively. The Stokes parameters are normalized to the average flux. The \(Q\) and \(U\) parameters are further related to the PD, PA, and polarized flux (PF) through the following expressions:
\begin{eqnarray}  
Q_{\mathrm{p}}(\phi) &=& \mathrm{PD}_{\mathrm{p}}(\phi) \, I_{\mathrm{p}}(\phi) \cos[2\chi(\phi)] = \mathrm{PF}_{\mathrm{p}}(\phi) \cos[2\chi(\phi)], \nonumber \\
U_{\mathrm{p}}(\phi) &=& \mathrm{PD}_{\mathrm{p}}(\phi) \, I_{\mathrm{p}}(\phi) \sin[2\chi(\phi)] = \mathrm{PF}_{\mathrm{p}}(\phi) \sin[2\chi(\phi)], \\
Q_{\mathrm{a}} &=& \mathrm{PD}_{\mathrm{a}} \, I_{\mathrm{a}} \cos(2\chi_{\mathrm{a}}) = \mathrm{PF}_{\mathrm{a}} \cos(2\chi_{\mathrm{a}}), \nonumber \\
U_{\mathrm{a}} &=& \mathrm{PD}_{\mathrm{a}} \, I_{\mathrm{a}} \sin(2\chi_{\mathrm{a}}) = \mathrm{PF}_{\mathrm{a}} \sin(2\chi_{\mathrm{a}}). \nonumber
\end{eqnarray} 
Here, PD\(_{\mathrm{p}}(\phi)\) and PD\(_{\mathrm{a}}\) represent the PDs of the pulsed and additional components, respectively; \(\chi(\phi)\) is the PA of the pulsed component, while \(\chi_{\mathrm{a}}\) denotes the PA of the additional component.

We fitted the observed phase-resolved Stokes parameters $I, Q$, and $U$, assuming uniform priors for most parameters. For the inclination angle $i_{\mathrm{p}}$ and magnetic obliquity $\theta$, flat priors were assumed for $\cos i_{\mathrm{p}}$ and $\cos \theta$. The likelihood function was constructed using $\chi^2$ statistics for $Q$ and $U$ following \citet{Doroshenko_etal_2023}. The parameter ranges were set as follows: $i_{\mathrm{p}} \in [0\degr, 180\degr]$, $\theta \in [0\degr, 90\degr]$, $\chi_{\mathrm{p}} \in [-90\degr, 90\degr]$, $\phi_{\mathrm{p}}/(2{\pi}) \in [0, 1]$, $\mathrm{PD}_{\mathrm{a}} \in [0, 0.3]$, $\mathrm{PD}_{\mathrm{p}} \in [0, 1]$, and $I_{\mathrm{a}} \leq \min[I(\phi)]$. The MCMC fitting yields posterior distributions for the RVM geometrical parameters and the unpulsed Stokes components ($I_{\rm a}$, $Q_{\rm a}$, and $U_{\rm a}$). From these unpulsed components, we derived the distributions of the polarization flux and angle for the constant component ($PF_{\rm c}$ and $\chi_{\rm c}$), which are displayed in Fig.~\ref{fig:corner_plot_rvm2}. To determine the phase-resolved properties of the pulsed component, we subtracted the unpulsed contribution from the observed Stokes parameters for each sample in the chain. This allowed us to derive the posterior distributions for the pulsed PA ($\mathrm{PA}_{\mathrm{p}}$) as shown in Figs.~\ref{fig:corner_PA1}--\ref{fig:corner_PA3}.
Figure~\ref{fig:two-comp-fit} shows the phase-resolved PAs of the pulsed components in three energy bands, overlaid with the best-fit RVM curves. The dashed lines indicate the inferred PA of the unpulsed component, which remains roughly the same across different energy bands. However, even with the inclusion of this additional component, significant deviations from the RVM predictions persist in a few phase bins.

%%%%%%%%%%%%%%%%%%%%%%%%%%%%%%%%%%%
\begin{figure*} %[h!]
\centering
\includegraphics[width=0.9\textwidth]{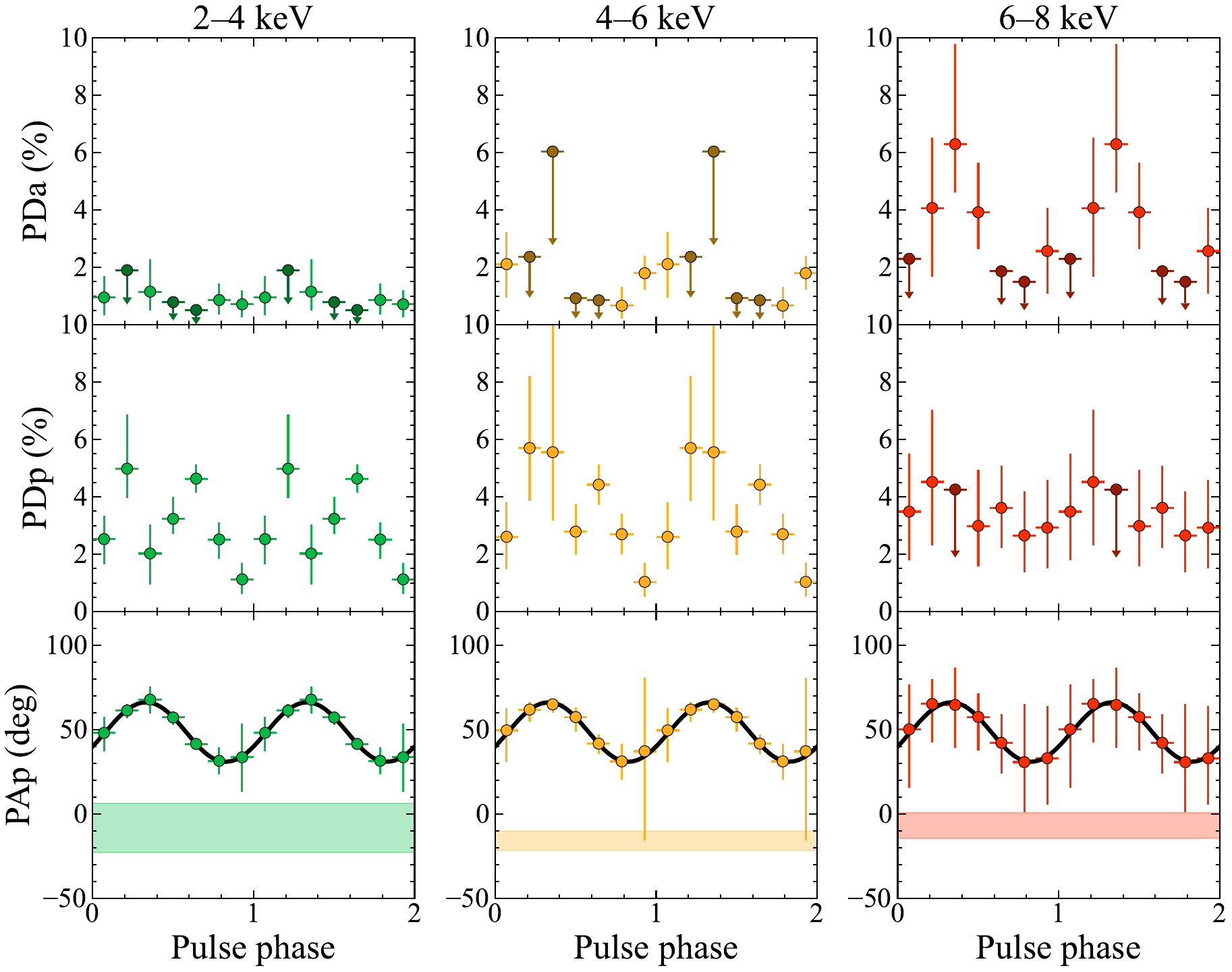}
\caption{Pulse phase-resolved polarization properties decomposed into pulsed and additional components. For each energy band, the top panels show the PDs (in unit of total flux) of the additional component (PDa), and the middle panels show the PDs of the pulsed component (PDp). When the PDs are detected with a confidence level below 68\% (1$\sigma$), the corresponding 1$\sigma$ upper limits are shown as downward arrows in darker colors.
%When the pulsed PD is not significantly detected at the 68\% confidence level, a corresponding upper limit is shown (solid darker markers).
The bottom panels display the PA of the pulsed component (PAp) versus pulse phase. The shaded horizontal bands indicate the phase-independent PAs of the additional component, while the black solid curves represent the best-fit RVM to the pulsed component.}
%Top panel: PD of the scattered (additional) component normalized by the total flux in each phase bin. Middle panel: PD of the pulsed component normalized by the total flux in each phase bin. \red{We use the upper limit at 68\% when the measured PD is not significant at this confidence level.} Bottom panel: PA evolution where crosses show the pulsed component and colored bands represent the constant PA of the scattered component. The black solid line shows the best-fit RVM model for the pulsed component.
%The color coding is the same as in Fig.~\ref{fig:phase_resolved}. }
\label{fig:phase_energy_new}
\end{figure*}
%%%%%%%%%%%%%%%%%%%%%%%%%%%%%%%%%%%

Notably, the energy dependence of PA is strongly phase-dependent, as illustrated in Fig.~\ref{fig:phase_resolved}. In most phase bins, the PAs are consistent across energy bands. However, in the phase intervals 0.286–0.429 and 0.429–0.572, the PA exhibits a clear energy dependence. To further illustrate this difference, we plot two phase intervals: 0.143--0.286 and 0.286--0.429.
%and 0.572--0.715. 
As shown in Fig.~\ref{fig:two_phase_energy}, these two intervals display qualitatively different PA evolution with energy: one shows a clear linear trend, while the other remains nearly constant. This suggests that the contribution of the additional component may be phase-dependent. It is interesting to note that the phase intervals 0.286–0.429 and 0.429–0.572 correspond to the main peak of the pulse profile.%This suggests that the contribution of the additional component may vary with pulse phase, implying it could also be phase-dependent.

%\begin{figure*} %[h!]
%\centering
%\includegraphics[width=0.95\textwidth]{fig/Tcpm_corner_RVM.pdf}
%\caption{Corner plots of the posterior distributions for the two-component RVM model with phase-variable scattered component. The parameters PD$_i$ ($i =$1--7) represent the polarized flux fraction of the scattered component normalized by the total flux in each of the seven phase bins. $\chi_{\rm a}$ denotes the constant PA of the scattered component. The color scheme is the same as in Fig.~\ref{fig:corner_plot_rvm}.}
%\label{fig:corner_plot_rvm2}
%\end{figure*}

The additional component is thought to originate from scattering in the disk wind or magnetospheric accretion flow. In such scenarios, the flux of scattered X-ray photons varies with pulse phase, resulting in an enhanced scattering contribution at specific viewing angles. Consequently, the Stokes parameters of this additional component exhibit phase-dependent variations.
To account for this possibility, we can revise Eq.~\eqref{eq:two-comp} by allowing the Stokes parameters of the additional component to vary with phase:
\begin{eqnarray}  
\label{eq:two-comp-new}
I(\phi) &=& I_{\mathrm{a}}(\phi) + I_{\mathrm{p}}(\phi), \nonumber \\
Q(\phi) &=& Q_{\mathrm{a}}(\phi) + Q_{\mathrm{p}}(\phi), \\
U(\phi) &=& U_{\mathrm{a}}(\phi) + U_{\mathrm{p}}(\phi). \nonumber
\end{eqnarray} 
%However, since the additional component still originates from the scattering scenarios discussed above, its PA is expected to remain approximately phase-independent, even if its polarized flux varies with pulse phase. 
The PA of the additional component depends on several factors, including the opening angle of the disk wind  and the angular distribution of the incident photons \citep{Nitindala2025}. For simplicity, we assume that the PA is constant and expressed as
\begin{align}
\label{eq:PA_2}
\chi_{\mathrm{a}} &= \frac{1}{2} \arctan\left( \frac{U_{\mathrm{a}}(\phi)}{Q_{\mathrm{a}}(\phi)} \right).
\end{align}

%%%%%%%%%%%%%%%%%%%%%%%%%%%%%%%%%%%%%%%%%%%%%%%
\begin{figure} %[h!]
\centering
\includegraphics[width=0.9\linewidth]{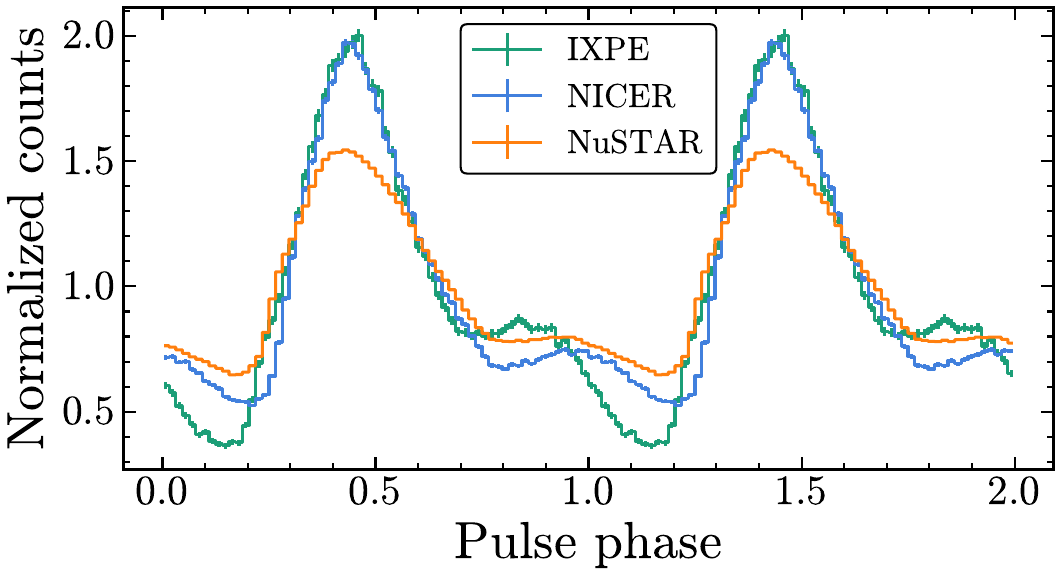}
\caption{Normalized pulse profiles of \ixpe\ (blue), \nicer\ (green), and \nustar\ (orange) in the same energy band (4--8 keV).}
\label{fig:pulse_profile_compar}
\end{figure}
%%%%%%%%%%%%%%%%%%%%%%%%%%%%%%%%%%%
\begin{figure} %[h!]
\centering
\includegraphics[width=0.45\textwidth]{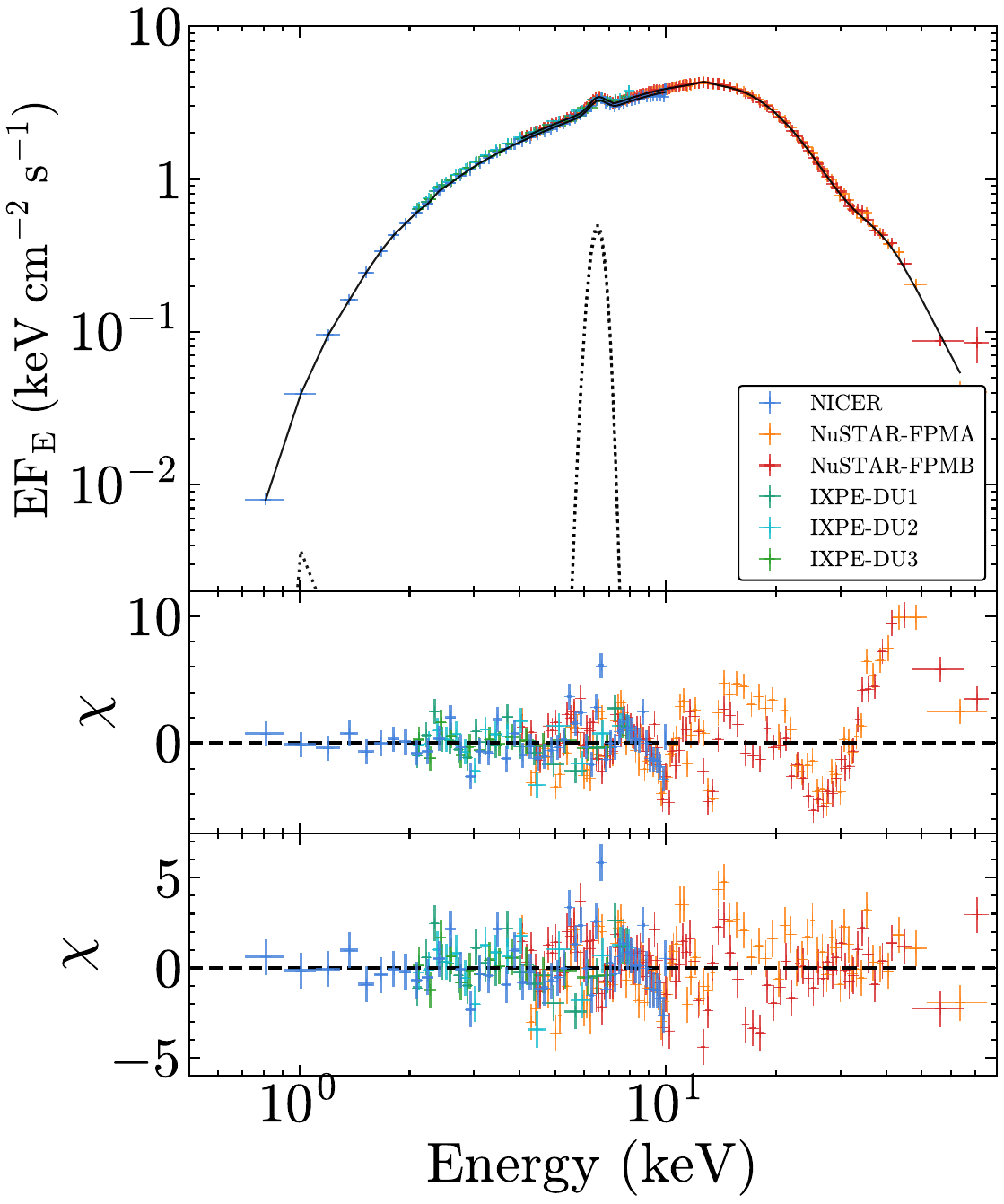}
\caption{Broadband spectral energy distribution of Cen X-3 with \ixpe, \nicer and \nustar.   Top panel: unfolded spectrum with the best-fit model (solid line). Middle panel: residuals showing a prominent absorption feature at $\sim$30 keV consistent with a CRSF. Bottom panel: residuals after including the CRSF component. The best-fit model is \texttt{constant*tbabs*pcfabs*(highecut*powerlaw*gabs*gabs + gauss + gauss)*gabs}.}
%The bottom panel displays the residuals when the CRSF (\texttt{gabs}) component is removed, highlighting the significance of the cyclotron feature around 30~keV.
\label{fig:spectra_fit_aver}
\end{figure}
%%%%%%%%%%%%%%%%%%%%%%%%%%%%%%%%%%%

Here we assume that only the polarized flux of the additional component varies with phase, while its PA \(\chi_{\mathrm{a}}\) remains fixed. Therefore, we treat \(\chi_{\mathrm{a}}\) as a global free parameter across all pulse phases, and allow \(Q_{\mathrm{a}}(\phi)\) to vary independently in each phase bin. The corresponding Stokes \(U\) parameter is then computed as:
\begin{align}
U_{\mathrm{a}}(\phi) = Q_{\mathrm{a}}(\phi) \tan (2\chi_{\mathrm{a}}).
\end{align}

In this analysis, we assume that the RVM parameters of the pulsed component are independent of energy. To mitigate the degeneracy between the pulsed and additional components and to reduce the number of free parameters, we fix the RVM geometry to the best-fit values obtained from the 2.0--4.0 keV band (see values in Table~\ref{tab:RVM fitting}). This choice is motivated by the fact that $i_{\rm p}$ from the 2.0--4.0 keV band are more consistent with the orbital inclination \citep{Ash1999}, and the magnetic obliquity $\theta$ agrees well with results from pulse profile modeling \citep{Kraus1996}.
For each energy band, we fit eight free parameters: seven parameters of \(Q_{\mathrm{a}}(\phi)\), corresponding to different pulse phase bins, and one parameter for the PA \(\chi_{\mathrm{a}}\) of the additional component. Across the three energy bands analyzed, the total number of free parameters amounts to 24. The fitting results are presented in Fig.~\ref{fig:phase_energy_new} and posterior distributions are plotted in Figs.~\ref{fig:corner_plot_rvm21}--\ref{fig:corner_plot_rvm23}.
The bottom panel of Fig.~\ref{fig:phase_energy_new} illustrates that the PA variations with pulse phase across different energy bands are well reproduced by a single set of RVM parameters, supporting the assumption of energy-independent geometry. The inferred PA of the additional component, \(\chi_{\mathrm{a}}\), remains approximately consistent across energy bands.
In this fitting, we assume that the PA of the additional component is constant with phase, while its polarized flux is allowed to vary with pulse phase.

\subsection{Spectral analysis}

To further investigate the origin of the observed energy dependence in the PA, We performed a broadband spectral analysis using \ixpe, \nicer, and \nustar observations. Since no simultaneous \nicer and \nustar observations were available during the \ixpe coverage, we selected archival observations corresponding to the same high state. The luminosities during the \ixpe, \nicer, and \nustar observations are comparable, being approximately $(2.3$--$2.4) \times 10^{37}\ \rm erg\ s^{-1}$ in the 2--8 keV energy band. In Fig.~\ref{fig:pulse_profile_compar}, we present the pulse profiles obtained with \ixpe, \nicer, and \nustar\ in the 4--8 keV band. The \ixpe\ pulse profile shows slight differences compared to that of \nicer. Since the observations are not simultaneous, such differences may reflect intrinsic temporal variability of the source, potentially including precessional effects of the neutron star. Furthermore, the pulse fraction measured by \nustar\ is significantly lower than that obtained from the simultaneous \nicer\ observation. This discrepancy may be attributable to dead-time effects in \nustar\ and warrants further investigation, which is beyond the scope of this paper.

For the spectra analysis, we adopted a commonly used phenomenological model for XRPs: \texttt{highecut*powerlaw}, where the \texttt{highecut} component is known for its sharp spectral drop at the cutoff energy, as widely discussed in previous works \citep{Kretschmar_cutoff, Kreykenbohm_cutoff, Coburn_cutoff}. To account for this sharp dip, we included a multiplicative \texttt{gabs} component. The centroid line energy is linked to the cutoff energy. Another \texttt{gabs} is included to account for the absorption structure around 2.2 keV in the \nicer spectrum, which is likely associated with the gold M-shell absorption edge. A \texttt{gaussian} component was added to model the iron fluorescence line, and an additional \texttt{gaussian} was introduced to address known calibration issues around $\sim$1~keV for \nicer. The interstellar absorption was modeled using \texttt{tbabs} while for intrinsic absorption, we further included the \texttt{pcfabs} component.% with the hydrogen column density fixed at $1.1\times10^{22}~\mathrm{cm^{-2}}$, based on the HI4PI Survey map \citep{HI4PI}, accessed via the HEASARC online tool.\footnote{\url{https://heasarc.gsfc.nasa.gov/cgi-bin/Tools/w3nh/w3nh.pl}} 

As shown in the bottom panel of Fig.~\ref{fig:spectra_fit_aver}, the spectrum exhibits a clear cyclotron resonance scattering feature (CRSF) around 30 keV, which we modeled with an additional \texttt{gabs} component. To correct for cross-calibration offsets between instruments, we applied the \texttt{constant} model. 
%which modifies the continuum using a multiplicative term of the form \( KE^{\Delta \Gamma} \).
Thus, our final spectral model is:
\begin{center}
\texttt{constant*tbabs*pcfabs*(highecut*powerlaw*gabs*gabs + gauss + gauss)*gabs}.
\end{center}

This model provides a satisfactory fit to the broadband spectrum over the 0.7--79~keV range. In this fitting, We let the instrument gain of \nicer and \ixpe as the free parameters to fit.
The best-fit parameter values are listed in Table~\ref{tab:joint_spec_fit}. 
The hydrogen column density is \(N_{\rm H} \approx 1.256\times10^{22}\,\mathrm{cm}^{-2}\), which is close to the value \(1.1\times10^{22}\,\mathrm{cm}^{-2}\) inferred from the HI4PI survey map \citep{HI4PI}.\footnote{\url{https://heasarc.gsfc.nasa.gov/cgi-bin/Tools/w3nh/w3nh.pl}}

\begin{figure*}[h!]
\centering
\includegraphics[width=0.85\textwidth]{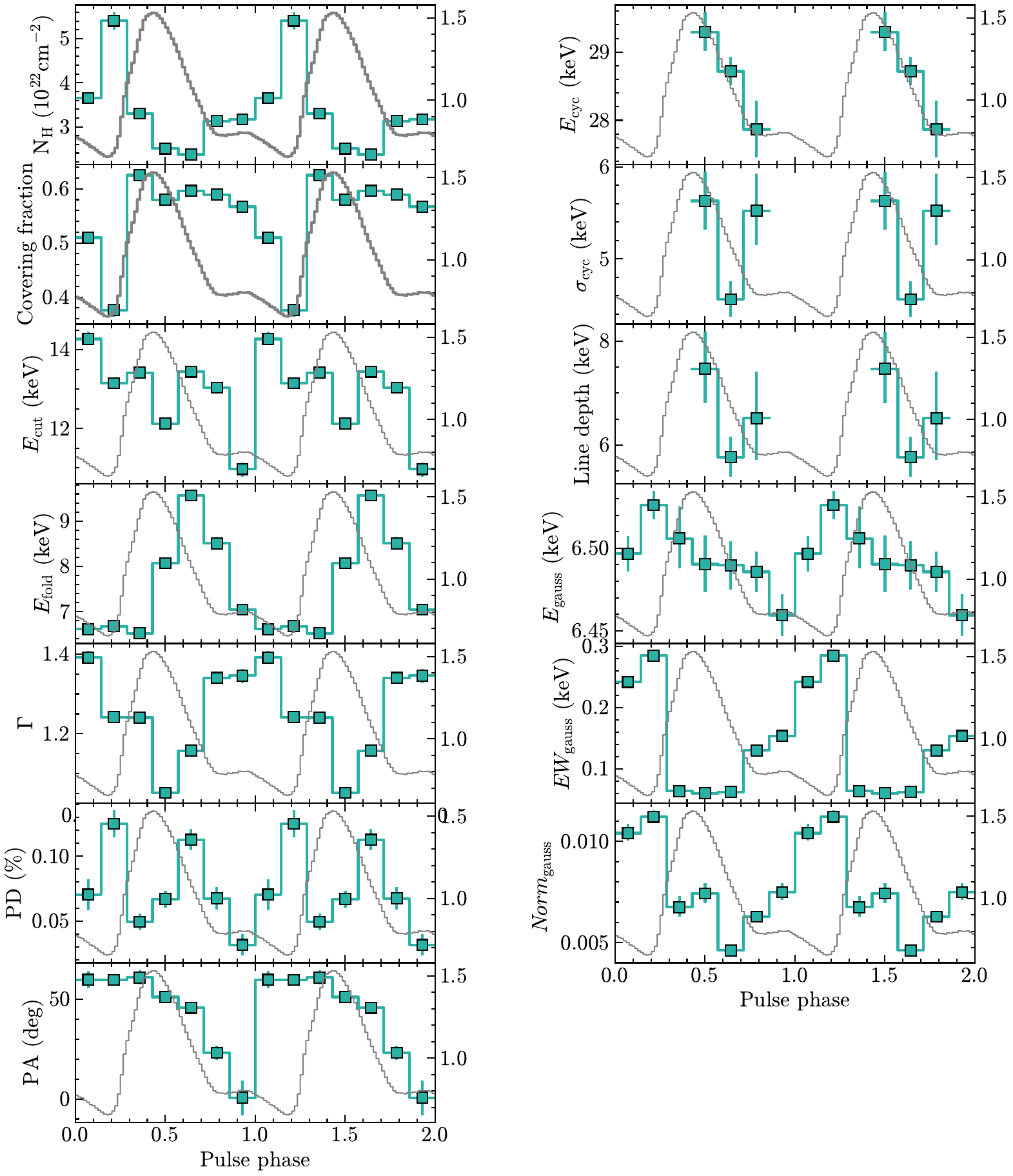}
\caption{Variation of the spectral parameters with pulse phase. Left axes exhibit the spectral parameters evolution with pulse phase, while the right axes show the pulse profiles of \nustar in 4--79 keV. The parameters shown are: $N_{\rm H}$ -- hydrogen column density from the \texttt{pcfabs} component ($\times 10^{22}$ cm$^{-2}$), representing intrinsic absorption; Covering fraction -- fraction of the source covered by absorbing material; $E_{\rm cut}$ -- cutoff energy in the \texttt{highecut} component (keV); $E_{\rm fold}$ -- e-folding energy in the \texttt{highecut} component (keV); $\Gamma$ -- photon index of the power-law component; PD -- polarization degree; PA -- polarization angle;$E_{\rm cyc}$ -- cyclotron resonance scattering feature (CRSF) centroid energy (keV); $\sigma_{\rm cyc}$ -- width of the CRSF line (keV); Line depth -- optical depth of the CRSF; $E_{\rm gauss}$ -- centroid energy of the iron fluorescence line (keV); $EW_{\rm gauss}$ -- equivalent width of the iron line (keV); ${\rm Norm}_{\rm gauss}$ -- normalization of the iron line component.}%Variation of the spectral parameters  with pulse phase. Left axes exhibit the spectra parameters from the model \texttt{crabcor*tbabs*pcfabs*(highecut*powerlaw*gabs + gauss + gauss)*gabs}, while the right axes show the pulse profiles of \textit{NuSTAR} in 4--79 keV.}
    \label{fig:phase_spectra}
\end{figure*}

We then performed a phase-resolved spectral analysis using the same model described above. For pulse phase intervals where the CRSF was not statistically significant, we omitted the additional \texttt{gabs} component. For the phase-resolved fits, we fixed the column density of \texttt{tbabs} to the phase-averaged best-fit value, \(N_{\rm H}=1.256\times10^{22}\,\mathrm{cm}^{-2}\), since interstellar absorption is not expected to vary with pulse phase. The best-fit parameters for the phase-resolved spectral analysis are listed in Table~\ref{tab:phase_resolved_joint_spec_fit} and are shown in Fig.~\ref{fig:phase_spectra}.
The parameters exhibit strong phase-dependent variability, as expected given the highly anisotropic emission reginon geometry of XRPs. %To examine potential degeneracies among spectral parameters, we performed a MCMC analysis using \textsc{xspec} and generated contour plots for representative parameter pairs. While some correlations between parameters are present, the overall variability with pulse phase remains robust and statistically significant.

Although most spectral parameters exhibit complex evolution with pulse phase, the photon index $\Gamma$ shows an anti-correlation with flux, while the CRSF line energy displays a clear positive correlation. Of particular interest are the hydrogen column density (\(N_{\rm H}\)) and the covering fraction from the \texttt{pcfabs} component, which we interpret as tracers of intervening wind material. Both parameters exhibit significant modulation with pulse phase, suggesting that the properties of the wind may also vary over the pulsation cycle.

\begin{table}%[ht]
\centering
\caption{Best-fit parameters for phase-averaged joint spectra of \ixpe, \nicer, and \nustar. }
\scalebox{0.9}{
\begin{tabular}{lcc}
\hline\hline
Model Component & Parameter & Value \\
\hline
TBabs & $N_{\rm H}$ ($10^{22} \ \rm cm^{-2}$) & $1.256^{+0.005}_{-0.015}$ \\ \hline
Pcfabs & $N_{\rm H}$ ($10^{22} \ \rm cm^{-2}$) & $2.88^{+0.06}_{-0.08}$ \\
& Covering fraction & $0.570^{+0.006}_{-0.004}$ \\ \hline
Highecut & $E_{\rm cut}$ (keV) & $12.71^{+0.03}_{-0.02}$ \\
& $E_{\rm fold}$ (keV) & $8.21\pm0.02$ \\ \hline
Powerlaw & $\Gamma$ & $1.221^{+0.004}_{-0.003}$ \\
& Normalization & $0.681^{+0.005}_{-0.003}$ \\ \hline
Gabs  & $E_{\rm line}$ (keV) & $2.21\pm0.01$ \\
& Sigma (keV) & $0.06\pm0.01$ \\
& Depth ($10^{-3}$) & $63^{+1}_{-3}$ \\ \hline
Gabs  & $E_{\rm line}$ (keV) & = $E_{\rm cut}$\\
& Sigma (keV) & $2.07^{+0.03}_{-0.04}$ \\
& Depth & $0.68^{+0.02}_{-0.01}$ \\ \hline
Gauss & $E_{\rm line}$ (keV) & $1.044^{+0.006}_{-0.007}$ \\
& Sigma (keV) & $0.056^{+0.006}_{-0.009}$ \\
& Normalization ($10^{-3}$) & $14^{+2}_{-1}$ \\ \hline
Gauss  & $E_{\rm line}$ (keV) & $6.493\pm0.005$ \\
& Sigma (keV) & $0.292^{+0.006}_{-0.005}$ \\
& Normalization ($10^{-3}$)  & $8.7\pm0.1$ \\ \hline
Gabs  & $E_{\rm line}$ (keV) & $30.3^{+0.2}_{-0.1}$ \\
& Sigma (keV) & $6.0\pm0.1$ \\
& Depth & $6.1^{+0.3}_{-0.2}$ \\
\hline
Constant & NICER & $1.00^{\rm fixed}$ \\
& NuSTAR (FPMA) & $1.047^{+0.001}_{-0.002}$ \\
& NuSTAR (FPMB) & $1.054^{+0.002}_{-0.001}$ \\
& IXPE (DU1) & $1.043^{+0.003}_{-0.004}$ \\
& IXPE (DU2) & $1.063^{+0.002}_{-0.004}$ \\
& IXPE (DU3) & $1.029^{+0.002}_{-0.003}$ \\ \hline
Gain & NICER Gain slope & $1.0177^{+0.0004}_{-0.0009}$ \\
& NICER Gain offset (keV) & $-0.035^{+0.003}_{-0.002}$ \\ 
 & IXPE (DU1) Gain slope & $1.010^{+0.001}_{-0.002}$ \\
& IXPE (DU1) Gain offset (keV) & $-0.043^{+0.007}_{-0.003}$ \\ 
 & IXPE (DU2) Gain slope & $1.006^{+0.001}_{-0.002}$ \\
& IXPE (DU2) Gain offset (keV) & $-0.022^{+0.007}_{-0.004}$ \\ 
 & IXPE (DU3) Gain slope & $1.008\pm0.001$ \\
& IXPE (DU3) Gain offset (keV) & $-0.016\pm0.004$ \\ \hline

& $\chi^2$/dof & $3235.8 / 3119$ \\
\hline
\end{tabular}
}
\label{tab:joint_spec_fit}
\end{table}

%%%%%%%%%%%%%%%%%%%%%%%%%%%%%%%%%%%
%%%%%%%%%%%%%%%%%%%%%%%%%%%%%%%%%%%

%
%最后做一些乱七八糟的讨论。
%首先就是重点讨论的是偏振的能量依赖性。PD现在没有模型可以解释，所以...但是PA在大部分源里面是很好的符合QED的预测，可以用RVM来建模。
%Vela x-1的结果是差90°，这个是由真空共振导致的。
%4U 1538是差70°，这个目前没有解释的很好。
%指出Cen X-3是如何变化的。是随相位依赖的，这表明不是用RX J0440和Swift J0243里面的额外的常数成分所解释的。
%相位分解谱表明：nh和covering fraction是随相位变化的。中子星的辐射照射到盘上，所驱动的盘风的性质是随脉冲相位而变化的。因为额外成分随相位不变的解释可能是不完全正确的。但是又很难去定量的评估如何随相位变化balabala>
%包括铁线的变化也可以一定程度说明风的位置有变化，所以相对于中子星的solid angle是存在变化的。
%最后在简单讨论一下cenx-3里面的回旋吸收线的情况。是对这个源回旋吸收线的简单更新。大约就是30keV.
%\begin{figure*}
%    \centering
   % \includegraphics[width=0.3\textwidth]{fig/Phase_0.00-0.125_polar_plot.pdf}
   % \includegraphics[width=0.3\textwidth]{fig/Phase_0.125-0.250_polar_plot.pdf}
   % \includegraphics[width=0.3\textwidth]{fig/Phase_0.250-0.375_polar_plot.pdf}
   % \includegraphics[width=0.3\textwidth]{fig/Phase_0.375-0.500_polar_plot.pdf}
   % \includegraphics[width=0.3\textwidth]{fig/Phase_0.500-0.625_polar_plot.pdf}
    %\includegraphics[width=0.3\textwidth]{fig/Phase_0.625-0.750_polar_plot.pdf}
    %\includegraphics[width=0.3\textwidth]{fig/Phase_0.750-0.875_polar_plot.pdf}
    %\includegraphics[width=0.3\textwidth]{fig/Phase_0.875-1.00_polar_plot.pdf}
    
    %\caption{Enter Caption}
%    \label{fig:enter-label}
%\end{figure*}

\section{Discussion and summary}
\label{sec:sec4}

\ixpe observations of XRPs have revealed both new insights and challenges in understanding these systems. The observed low PDs, which fall significantly below theoretical predictions, demand a re-examination of current models, particularly with respect to potential depolarization mechanisms. Further theoretical work will help elucidate the underlying mechanisms responsible for the observed PDs and their energy dependence.
Despite this, the variations in PA with pulse phase in most XRPs continue to be well described by the RVM, consistent with the predictions of vacuum birefringence. In this framework, the observed PA tracks the direction of the local magnetic field, allowing for the determination of the geometric parameters of XRPs. Since PA is expected to follow magnetic field geometry, it should show little to no dependence on photon energy for the radiation within the adiabatic radius.

%Thanks to \ixpe's energy resolution, we can investigate the energy dependence of polarization properties in XRPs. 
\citet{Tsygankov_etal_2022_cenx-3} performed a detailed polarimetric study of Cen X-3, including phase-averaged, phase-resolved, and phase-averaged energy-resolved analysis. In the present work, we extend this investigation by examining the phase-resolved energy dependence of the polarization. By dividing the data into three equal energy bands, we perform a phase-resolved analysis within each band.
Our results show that, while the phase-averaged PA exhibits only a weak dependence on energy, a few phase bins display a pronounced PA energy dependence. Moreover, the pattern of PA variation with energy differs across phase bins, as illustrated in Fig.~\ref{fig:two_phase_energy}. This strong, phase-dependent energy behavior of the PA is inconsistent with the predictions of vacuum birefringence.

\begin{table*}
%\begin{sidewaystable}
\centering
\footnotesize 
\caption{Best-fit parameters for the phase-resolved \ixpe (2--8 keV), \nicer (0.7--10 keV), and \nustar (4--79 keV) spectra.} 
%We note that the large cross-calibration constants of \nustar in some pulse phases are likely due to dead-time effects.}
\begin{tabular}{lccccccc}
\hline\hline 
 & \multicolumn{7}{c}{Phase} \\
\cline{2-8} 
Parameter & 0.000–0.143 & 0.143–0.286 & 0.286–0.429 & 0.429–0.572 & 0.572–0.715 & 0.715–0.858 & 0.858–1.000 \\
\hline
 $N_{\rm H}$ ($10^{22}\ \rm cm^{-2}$) & $3.7\pm0.1$ & $5.4\pm0.2$ & $3.30\pm0.08$ & $2.51\pm0.07$ & $2.37\pm0.06$ & $3.13^{+0.09}_{-0.08}$ & $3.2\pm0.1$ \\
 Covering fraction & $0.51\pm0.01$ & $0.38\pm0.01$ & $0.625\pm0.006$ & $0.580\pm0.006$ & $0.596\pm0.006$ & $0.589\pm0.006$ & $0.57\pm0.01$ \\
 $E_{\rm cut}$ (keV) & $14.3\pm0.2$ & $13.2\pm0.1$ & $13.4\pm0.1$ & $12.1\pm0.1$ & $13.44\pm0.06$ & $13.0^{+0.2}_{-0.1}$ & $11.0\pm0.2$ \\
 $E_{\rm fold}$ (keV) & $6.6\pm0.1$ & $6.68\pm0.08$ & $6.52^{+0.07}_{-0.06}$ & $8.07\pm0.09$ & $9.6\pm0.1$ & $8.5\pm0.1$ & $7.04^{+0.08}_{-0.09}$ \\
 $\Gamma$ & $1.39\pm0.02$ & $1.24\pm0.01$ & $1.24\pm0.01$ & $1.052^{+0.007}_{-0.008}$ & $1.158\pm0.004$ & $1.340\pm0.009$ & $1.35\pm0.02$ \\
 Power-law norm & $0.58^{+0.02}_{-0.01}$ & $0.40\pm0.01$ & $1.09\pm0.02$ & $0.89\pm0.01$ & $0.648\pm0.005$ & $0.592\pm0.008$ & $0.60\pm0.02$ \\
 $E_{\rm line}$ (keV, Gauss) & $6.50\pm0.01$ & $6.526\pm0.009$ & $6.51\pm0.02$ & $6.49\pm0.02$ & $6.49\pm0.01$ & $6.49\pm0.01$ & $6.46\pm0.01$ \\
 $\sigma$ (keV, Gauss) & $0.35\pm0.02$ & $0.34\pm0.01$ & $0.23\pm0.03$ & $0.23\pm0.02$ & $0.17\pm0.02$ & $0.25\pm0.02$ & $0.30\pm0.02$ \\
 Gauss norm ($10^{-3}$) & $10.4^{+0.5}_{-0.4}$ & $11.2\pm0.3$ & $6.8\pm0.5$ & $7.4\pm0.5$ & $4.6\pm0.3$ & $6.3\pm0.3$ & $7.5\pm0.4$ \\
 EW (keV) & $0.242\pm0.008$ & $0.285\pm0.008$ & $0.064\pm0.004$ & $0.060\pm0.004$ & $0.062\pm0.004$ & $0.130^{+0.006}_{-0.005}$ & $0.154\pm0.007$ \\

 $E_{\rm cyc}$ (keV) & – & – & – & $29.3\pm0.3$ & $28.7\pm0.2$ & $27.9\pm0.4$ & – \\
 $\sigma_{\rm cyc}$ (keV) & – & – & – & $5.6\pm0.3$ & $4.6\pm0.2$ & $5.5\pm0.4$ & – \\
 $\rm Depth_{\rm cyc}$ & – & – & – & $7.5\pm0.7$ & $5.8\pm0.4$ & $6.5^{+0.9}_{-0.8}$ & – \\
 Constant (NICER) & $1.00^{\rm fixed}$ & $1.00^{\rm fixed}$ & $1.00^{\rm fixed}$ & $1.00^{\rm fixed}$ & $1.00^{\rm fixed}$ & $1.00^{\rm fixed}$ & $1.00^{\rm fixed}$ \\
 Constant (FPMA) & $1.152\pm0.005$ & $1.400\pm0.006$ & $0.999\pm0.003$ & $0.870\pm0.002$ & $1.040\pm0.003$ & $1.230\pm0.005$ & $1.214\pm0.005$ \\
 Constant (FPMB) & $1.159\pm0.005$ & $1.369\pm0.006$ & $1.001\pm0.003$ & $0.886\pm0.002$ & $1.056\pm0.003$ & $1.231\pm0.005$ & $1.218\pm0.005$ \\
  Constant (DU1) & $0.837\pm0.004$ & $1.224\pm0.006$ & $1.039\pm0.003$ & $1.090\pm0.003$ & $1.075\pm0.003$ & $1.251\pm0.005$ & $1.115\pm0.005$ \\
   Constant (DU2) & $0.875\pm0.005$ & $1.220\pm0.006$ & $1.040\pm0.003$ & $1.134\pm0.003$ & $1.133\pm0.004$ & $1.292\pm0.006$ & $1.158\pm0.005$ \\
    Constant (DU3) & $0.835\pm0.005$ & $1.233\pm0.006$ & $1.042\pm0.003$ & $1.086\pm0.003$ & $1.081\pm0.004$ & $1.256\pm0.005$ & $1.103\pm0.005$ \\ \hline
 $\chi^2$/dof & $2659.9/2361$ & $2750.0/2391$ & $3007.2/2616$ & $2953.0/2680$ & $2907.1/2670$ & $2601.7/2419$ & $2509.3/2357$ \\
\hline 
\end{tabular}
\label{tab:phase_resolved_joint_spec_fit}

%\end{sidewaystable}
\end{table*}

However, mode conversion between the O- and X- modes at the vacuum resonance can introduce a $90\degr$ shift in PA. \ixpe  observations of Vela X-1 \citep{Forsblom2025} revealed exactly such a $90\degr$ PA swing between low- and high-energy bands in phase-averaged data. Phase-resolved analysis detected significant polarization in only one phase interval at low energy; comparison with the corresponding high-energy interval again showed a $90\degr$ offset. The observed signatures for this source are naturally explained by vacuum-resonance–driven mode conversion. In Cen X-3, however, the PA evolution with energy departs from the exact $90\degr$ flip predicted for vacuum-resonance mode conversion. Instead, the PA in the phase range 0.286--0.429 shows a distinct linear trend with energy as shown in  Fig.~\ref{fig:two_phase_energy}, indicating that this mechanism alone cannot account for the observed behavior. A phase-averaged investigation by \citet{Loktev25} uncovered a $\sim 70\degr$ PA shift between low- and high-energy bands for \mbox{4U 1538$-$52}. Moreover, phase-resolved data present an even more intricate picture: pulse-phase–dependent PA offsets diverge from the $90\degr$ prediction, indicating that vacuum-resonance mode conversion cannot by itself explain the observed polarimetric behavior of this source.

Although the energy dependence of polarization remains insufficiently explored in many XRPs, some sources already exhibit pronounced epoch-to-epoch variations in their PA profiles as a function of pulse phase—for example, \mbox{RX~J0440.9+4431}/\mbox{LS V +44 17} \citep{Doroshenko_etal_2023,Zhao_RXJ0440} and \mbox{Swift~J0243.6+6124} \citep{SwiftJ0243_Poutanen}. In both cases, the PA varies markedly between observations. Introducing an additional, non-pulsating polarized component allows these disparate PA patterns to be reconciled with a single set of RVM parameters. This non-pulsating polarized component could plausibly arise from scattering in the disk wind or within the magnetospheric accretion flow.
Fig.~\ref{fig:two-comp-fit} shows the two-component RVM fit, which assumes a phase-independent polarized flux and PA for the additional component. Although introducing the constant polarized component, a few data points still deviate from the model curve, indicating that even this framework does not fully capture the observed behavior. As shown in Figs.~\ref{fig:phase_energy} and~\ref{fig:phase_resolved}, the strongest energy dependence of the PA occurs in the phase interval 0.286–0.572, corresponding to the peak of the pulse profile. This indicates that the polarization properties of this additional component may also vary with phase. We therefore relax the assumption of constant polarized flux and allow it to vary with pulse phase by considering that the observed polarized signal depends on the phase-dependent illuminating flux. In addition, variations in the disk wind properties may also lead to changes in the additional scattered component across pulse phase \citep{Nitindala2025}. This naturally leads to the expectation that the additional scattering-induced polarized flux should modulate with phase. 
As shown in Fig.~\ref{fig:phase_energy_new}, allowing the polarized flux to vary enables the PA of the pulsating component to be well reproduced using a single set of RVM parameters. In this fit, since the polarized flux is no longer fixed across phases, the four RVM parameters do not converge to meaningful values when left free. We therefore fix them to the best-fit values obtained from the 2–4~keV band. As evident from the fit, the polarized flux of the additional component varies with pulse phase. The PA of this component across different energy bands is broadly consistent. As discussed in \citet{Doroshenko_etal_2023}, this additional component may originate from scattering in the disk wind. Since the wind is located well beyond the adiabatic radius, the vacuum birefringence predictions can still hold true when this contribution is taken into account. If the additional component originates from scattering in the disk wind, its PA can naturally be associated with the disk axis, while optical polarization from scattering in the disk may provide an independent measure of the disk orientation. Therefore, optical polarization measurements would offer a valuable way to evaluate this interpretation.

We also performed a broad-band spectral analysis using \ixpe, \nicer~ and \nustar~ data. In this analysis, we employed the \texttt{pcfabs} model to trace the properties of the disk wind. A phase-resolved spectral analysis reveals that both the column density ($N_{\rm H}$) and covering fraction inferred from \texttt{pcfabs} exhibit significant modulation with pulse phase, as shown in Fig.~\ref{fig:phase_spectra}. %This modulation may be caused by changes in magnetic pressure over the pulse cycle.
The variation in covering fraction suggests that the scattered flux also varies with pulse phase, rather than remaining constant. In addition, a reflected Fe line is clearly present in the spectra and exhibits pronounced variations with pulse phase. These spectral results indicate that the disk wind properties vary with pulse phase, providing a natural explanation for the observed phase-dependent variations in the polarized flux of the additional component.

However, it should be noted that the polarization properties of the additional component depend on several factors, including the geometry of the scattering region and the beaming pattern of the pulsar emission. The response of the scattered component to the illuminating flux should be examined in greater detail -- for example, through dedicated Monte Carlo simulations.
In this work, due to limited photon statistics and for simplicity, we allow only the polarized flux of the additional component to vary freely. We note that the polarized flux of this component can reach up to approximately 6\% (normalized by the total flux). Assuming an inclination of $i = 70\degr$ and adopting the analytical expression ${\rm PD} = \sin^2 i / (3 - \cos^2 i)$ \citep{Sunyaev&Titarchuk}, the expected PD of the scattered component is about 30\% (self-normalized) and it is a weak function of inclination and the wind opening angle \citep{Nitindala2025}. This implies that the scattered component would contribute nearly 20\% of the total flux, which is a considerable fraction. Further refinements to this picture may emerge as we better understand the scattered component's actual flux contribution and polarization properties. The observed energy-dependent polarization behavior provides a useful context for developing theoretical models that incorporate more complex scattering geometries or additional polarization mechanisms. In addition, future missions such as the enhanced X-ray Timing and Polarimetry mission \citep[eXTP, ][]{eXTP,eXTP_WG3} will offer greatly improved polarimetric statistics, enabling more robust tests of these scenarios.

%%%%%%%%%%%%%%%%%%%%%%%%%%%%%%%%%%%%%%%%%%%

\begin{acknowledgements}
We thank the anonymous referee for the constructive comments that helped improve the manuscript. Financial support for this work is provided by the National Key R\&D Program of China (2021YFA0718500). We also acknowledge funding from the National Natural Science Foundation of China (NSFC) under grant numbers 12122306, 12333007, and U2038102. We acknowledge support from the China's  Space Origins Exploration Program. This research was supported by the International Space Science Institute (ISSI) in Bern, through International Team project 25-657 'Polarimetric Insights into Extreme Magnetism’. SST and JP acknowledge support by the Research Council of Finland, the Centre of Excellence in Neutron-Star Physics (project 374064).
\end{acknowledgements}
\bibliographystyle{yahapj}

\bibliography{ref}
\begin{appendix}
\section{Posterior distributions of two-component RVM}

%%%%%%%%%%%%%%%%%%%%%%%%%%%%%%%%%%%

\begin{figure*} %[h!]
\centering
\includegraphics[width=0.95\textwidth]{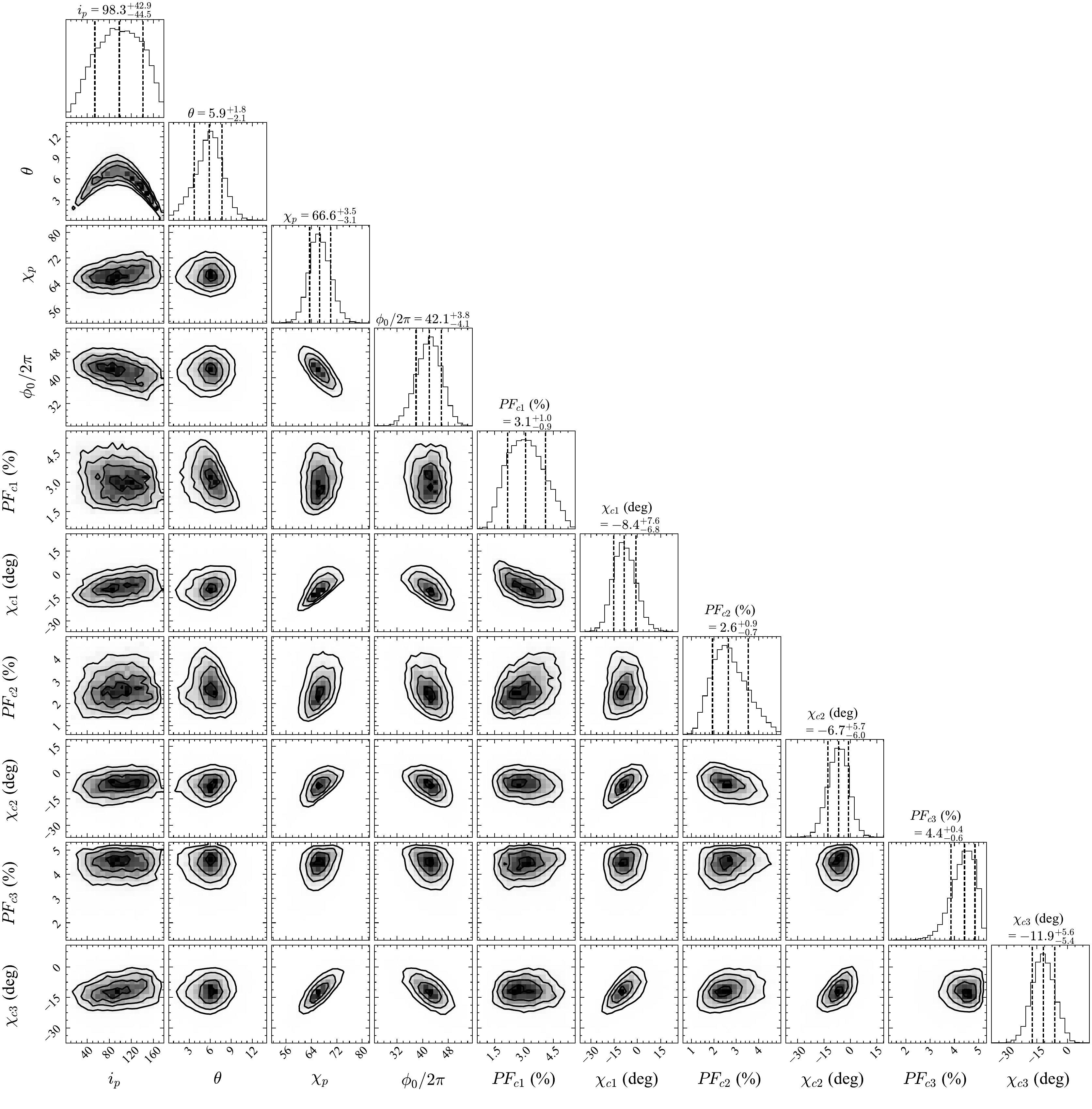}
\caption{Corner plots of the posterior distributions for the two-component RVM model with phase-independent additional component. $PF_{\mathrm c \ i}$, i = 1, 2, 3 are the polarized flux of the phase-independent additional component in unit of averaged flux for three energy bands (2--4, 4--6 and 6--8 keV). $\chi_{\mathrm c \ i}$ are its PA.}
\label{fig:corner_plot_rvm2}
\end{figure*}

\begin{figure*} %[h!]
\centering
\includegraphics[width=0.95\textwidth]{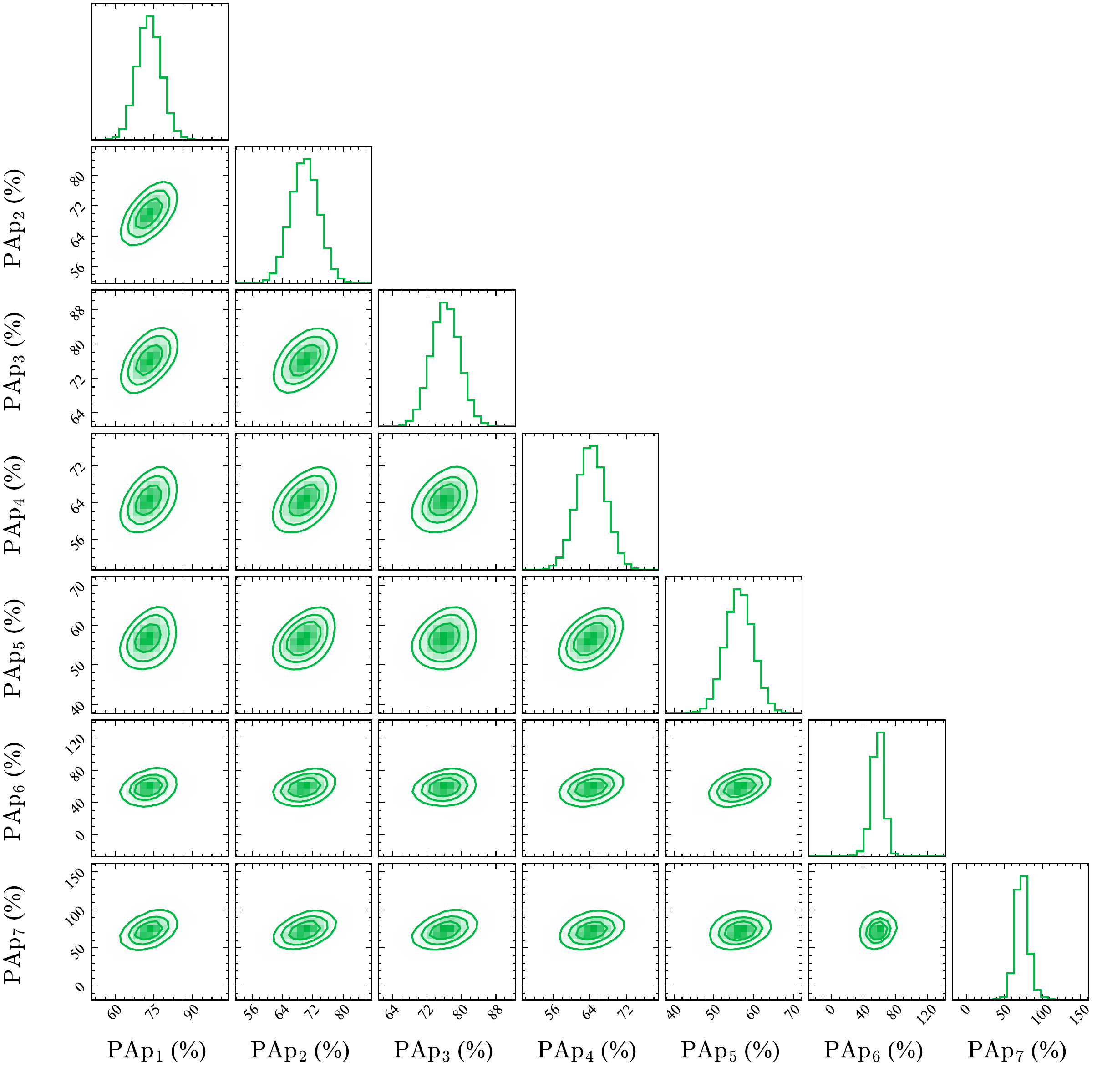}
\caption{Corner plots of the posterior distributions for PA of pulsed component in 2--4 keV energy band. The parameters PAp$_i$ ($i =$1--7) represent the PA of the pulsed component in each of the seven phase bins.}
\label{fig:corner_PA1}
\end{figure*}
\begin{figure*} %[h!]
\centering
\includegraphics[width=0.95\textwidth]{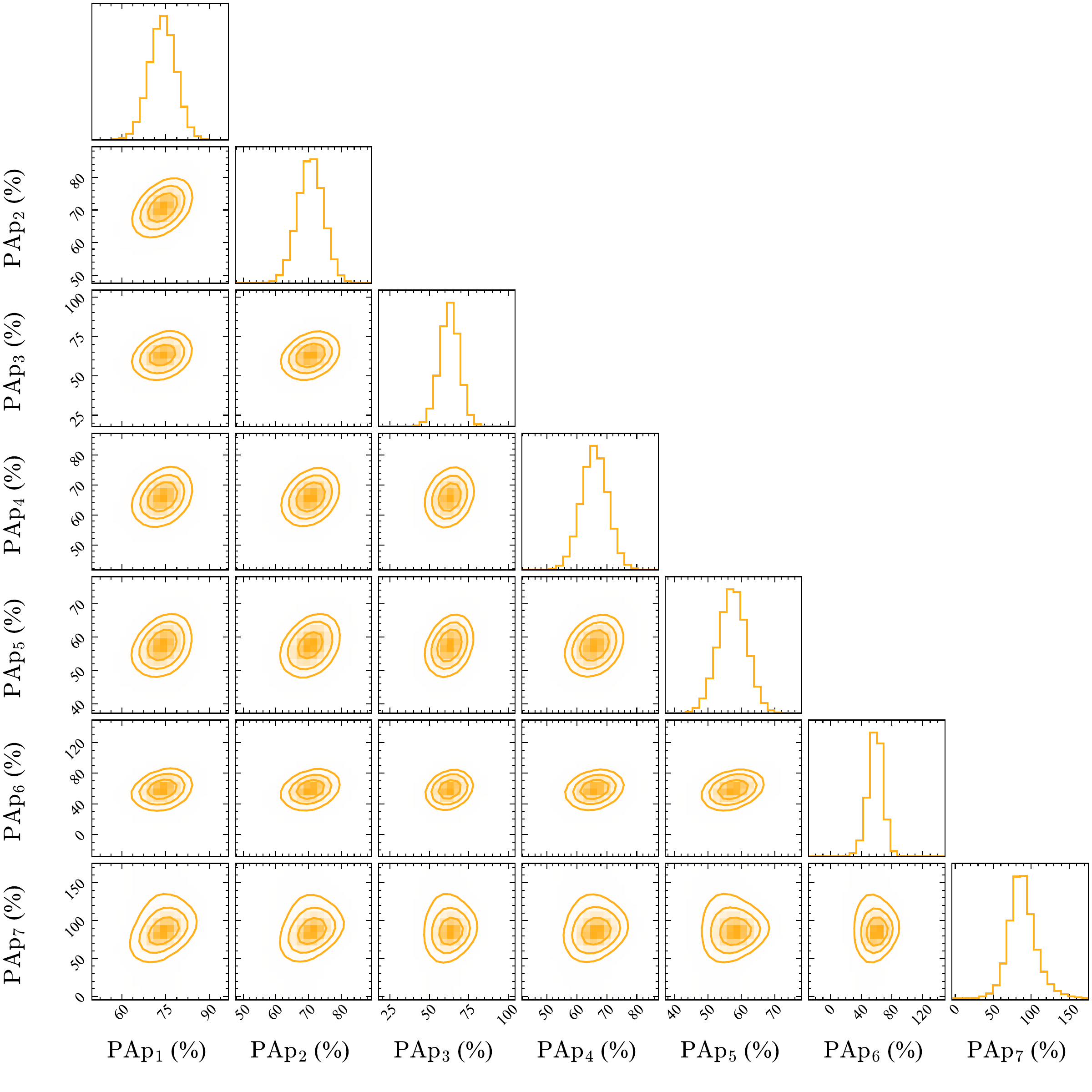}
\caption{Same as Fig.~\ref{fig:corner_PA1}, but for 4--6 keV energy band.}
\label{fig:corner_PA2}
\end{figure*}
\begin{figure*} %[h!]
\centering
\includegraphics[width=0.95\textwidth]{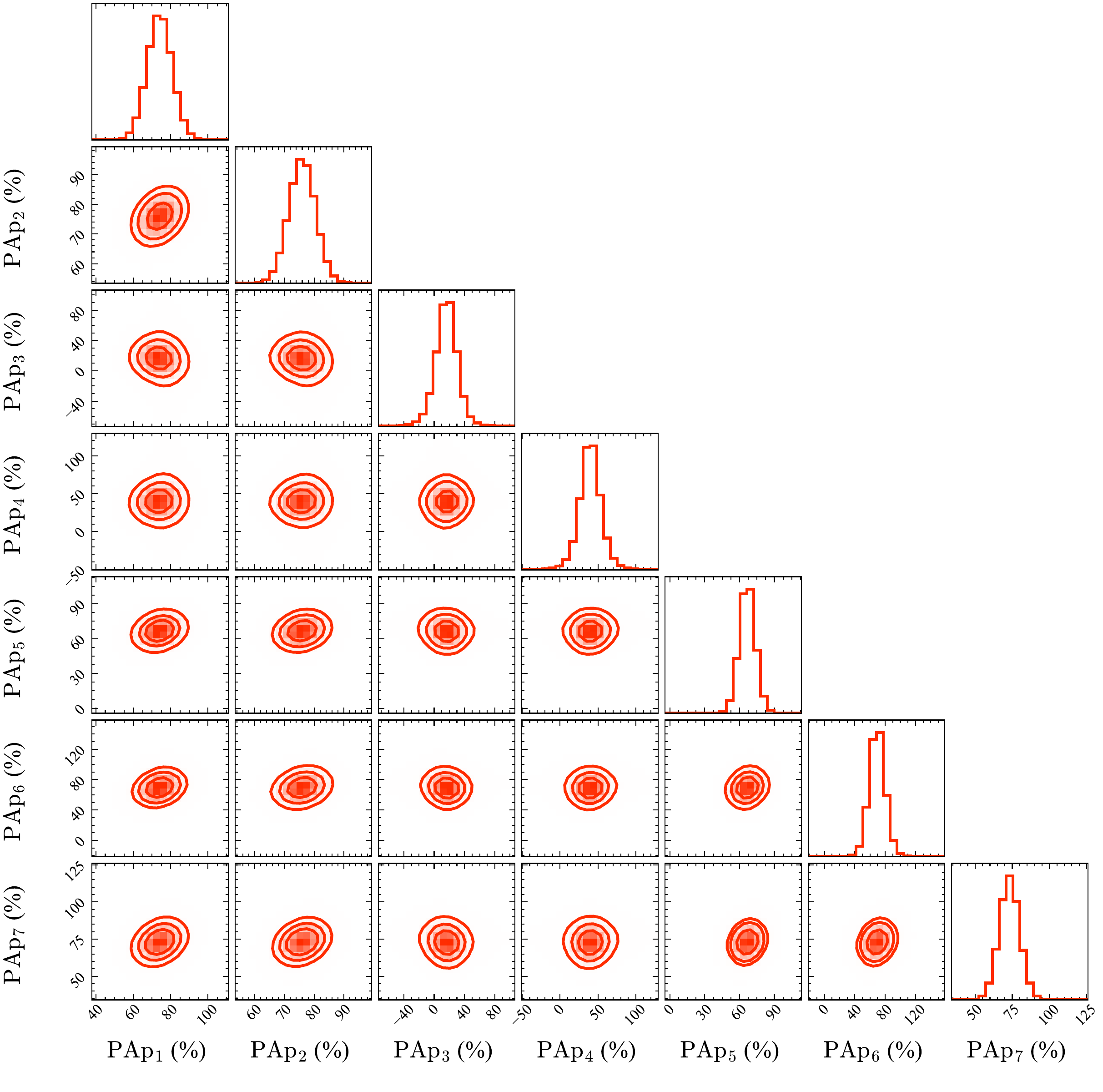}
\caption{Same as Fig.~\ref{fig:corner_PA1}, but for 6--8 keV energy band.}
\label{fig:corner_PA3}
\end{figure*}

\begin{figure*} %[h!]
\centering
\includegraphics[width=0.95\textwidth]{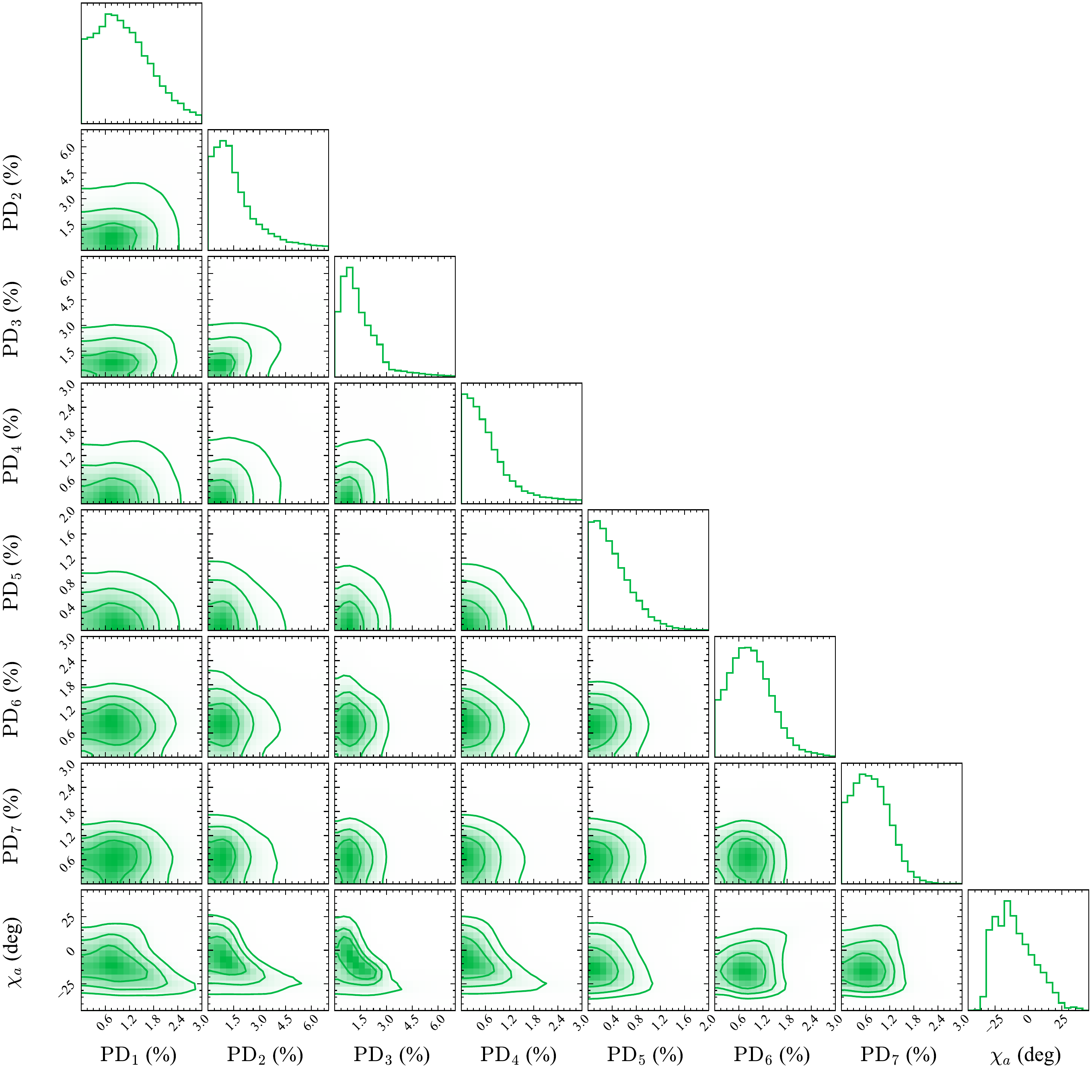}
\caption{Corner plots of the posterior distributions for the two-component RVM model with phase-variable additional component for 2--4 keV band. The parameters PD$_i$ ($i =$1--7) represent the polarized flux fraction of the scattered component normalized by the total flux in each of the seven phase bins. $\chi_{\rm a}$ denotes the constant PA of the scattered component.}
\label{fig:corner_plot_rvm21}
\end{figure*}
\begin{figure*} %[h!]
\centering
\includegraphics[width=0.95\textwidth]{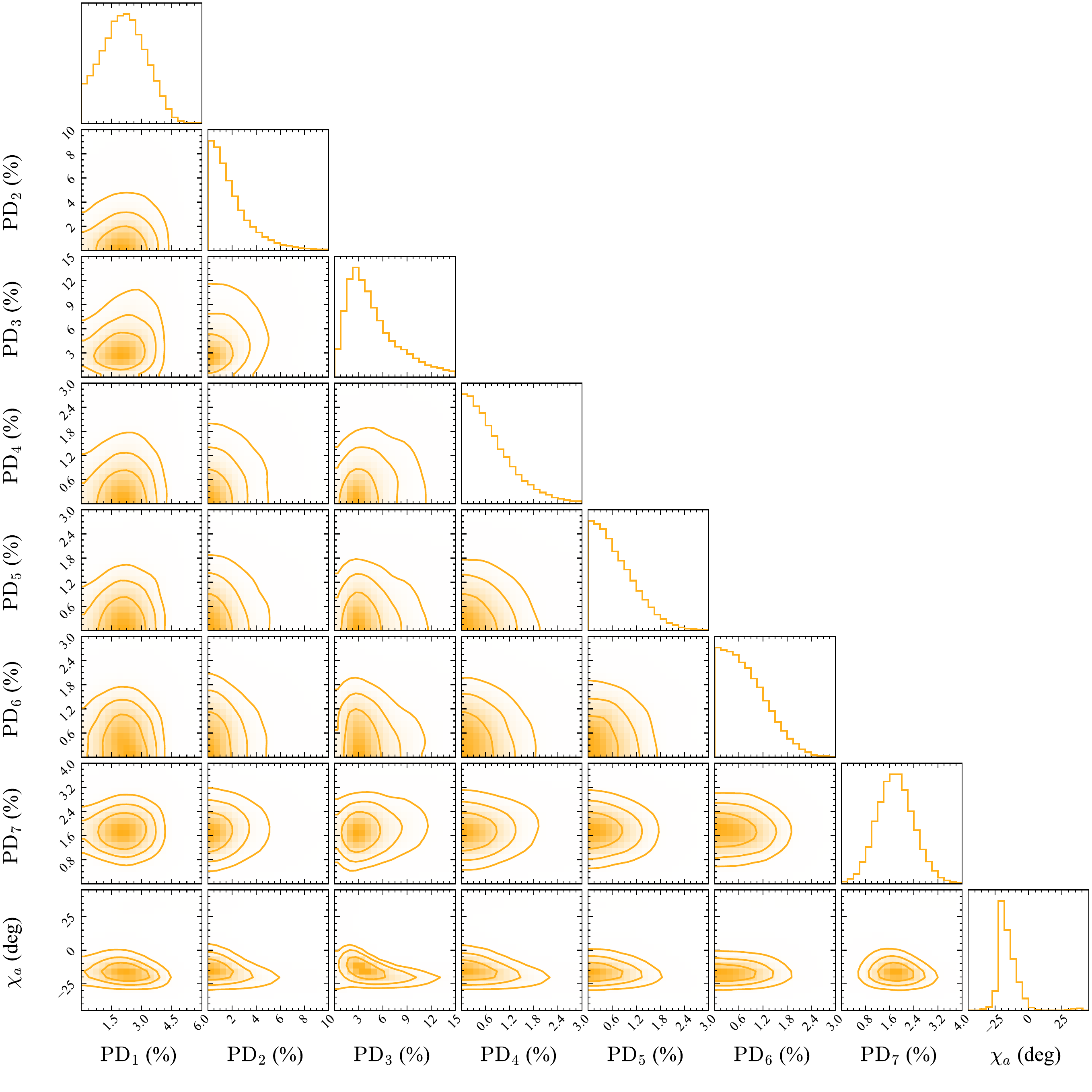}
\caption{Same as Fig.~\ref{fig:corner_plot_rvm21}, but for 4--6 keV band.}
\label{fig:corner_plot_rvm22}
\end{figure*}
\begin{figure*} %[h!]
\centering
\includegraphics[width=0.95\textwidth]{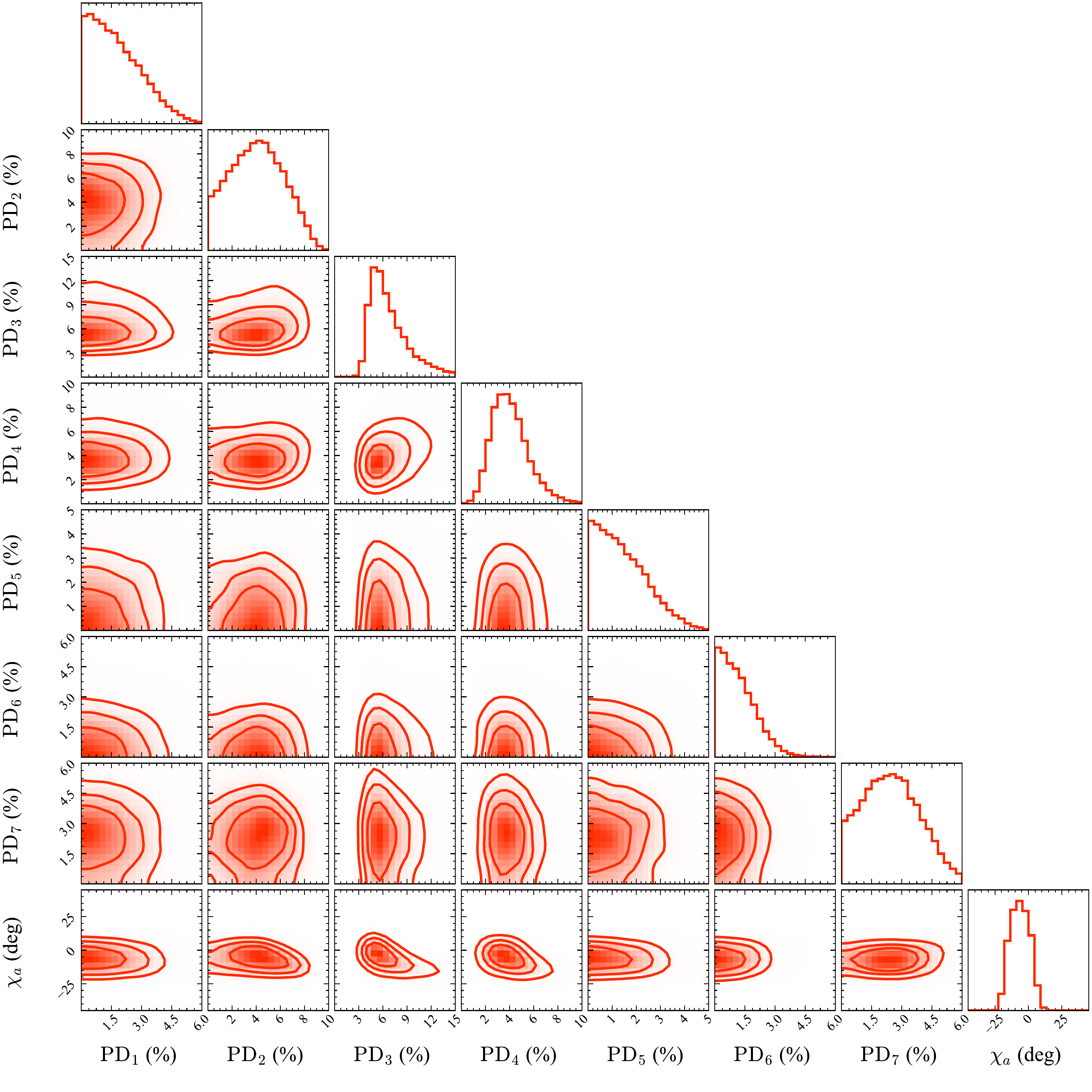}
\caption{Same as Fig.~\ref{fig:corner_plot_rvm21}, but for 6--8 keV band.}
\label{fig:corner_plot_rvm23}
\end{figure*}
\end{appendix}
\end{document}